\documentclass[12pt]{article}
\usepackage{amsfonts}
\usepackage{amsmath}
\usepackage{amssymb}
\usepackage{amsthm}
\usepackage{accents}
\usepackage{youngtab}
\usepackage{braket}
\usepackage{graphicx}
\usepackage[svgnames]{xcolor}
\usepackage{ytableau}
\usepackage{here}
\usepackage{comment}
\usepackage{tikz}
\usepackage{feynmf}
\usepackage{hyperref}
\usepackage{simplewick}

\newcommand{\ba}{\begin{eqnarray}}
\newcommand{\ea}{\end{eqnarray}}
\newcommand{\nn}{\nonumber}
\newcommand{\rrangle}{\rangle}

\newcommand{\g}{\mathfrak{g}}

\newcommand{\cW}{\mathcal{W}}

\newcommand{\cG}{\mathcal{G}}
\newcommand{\cY}{\mathcal{Y}}
\newcommand{\lt}{\left}
\newcommand{\rt}{\right}
\newcommand{\cO}{\mathcal{O}}
\newcommand{\Zvec}{\mathcal{Z}_{\text{vec}}}
\newcommand{\Zbf}{\mathcal{Z}_{\text{bifund}}}
\newcommand{\hg}{\hat{\gamma}}
\def\Zv{\mathcal{Z}_\text{vect.}}
\def\Zbf{\mathcal{Z}_\text{bfd.}}
\def\bZbf{\bar{\mathcal{Z}}_\text{bfd.}}
\def\Zf{\mathcal{Z}_\text{fund.}}
\def\Zaf{\mathcal{Z}_\text{a.f.}}
\def\Zinst{\mathcal{Z}_\text{inst.}}
\def\ZCS{\mathcal{Z}_\text{CS}}
\newcommand{\superp}[2]{\genfrac{}{}{0pt}{}{#1}{#2}}

 \def\d{\delta}

 \def\a{\alpha}
 \def\b{\beta}
 \def\g{\gamma}
 \def\d{\delta}
 \def\e{\epsilon}

 \def\k{\kappa}
 \def\l{\lambda}

 \def\s{\sigma}
 \def\t{\tau}

 \def\D{\Delta}

\def\CG{{\mathcal{G}}}

\def\CN{{\mathcal{N}}}

\def\CS{{\mathcal{S}}}
\def\CT{{\mathcal{T}}}

\def\CX{{\mathcal{X}}}

\def\la{\left\langle}
\def\ra{\right\rangle}

\def\bc{{\bar{c}}}

\def\implies{\quad\Rightarrow\quad}

\def\vphi{\varphi}

\def\Zv{\mathcal{Z}_\text{vect.}}
\def\Zbf{\mathcal{Z}_\text{bfd.}}
\def\bZbf{\bar{\mathcal{Z}}_\text{bfd.}}
\def\Zf{\mathcal{Z}_\text{fund.}}
\def\Zaf{\mathcal{Z}_\text{a.f.}}
\def\Zinst{\mathcal{Z}_\text{inst.}}
\def\ZCS{\mathcal{Z}_\text{CS}}

\def\aY{|\vec{v},\vec{\lambda}\rangle\rangle}
\def\baY{\langle\langle\vec{v},\vec{\lambda}|}

\def\qf{\mathfrak{q}}

\def\vac{\emptyset}

\def\hg{\hat\gamma}

\def\bd{\bar d}

\def\bt{\bar\tau}

 \DeclareMathOperator*{\Res}{Res}

\makeatletter
\newcommand{\douwidehat}[2]{%
  \sbox0{$\m@th#1\widehat{\hphantom{#2}}$}%
  \sbox2{$\m@th#1x$}
  \sbox4{$\m@th#1#2$}
  \dimen0=\ht0
  \advance\dimen0 -.8\ht2
  \dimen2=\dp4
  \rlap{%
    \raisebox{\dimexpr\dimen0-\dimen2}{%
      \scalebox{1}[-1]{\box0}%
    }%
  }%
  {#2}%
}
\makeatother

\makeatletter
\@addtoreset{equation}{section}

\makeatother

\begin{document}
\begin{titlepage}
\vspace*{-2cm}
	\begin{flushright}
		KIAS-Q17026\\
		UT-17-30
	\end{flushright}
	
	\vskip 12mm
	
	\begin{center}
		{\Large\bf Reflection states in Ding-Iohara-Miki algebra and brane-web for D-type quiver}
	\end{center}
	\vfill
		\begin{center}
			{\Large J.-E. Bourgine$^\dagger$, M. Fukuda$^\ast$,  Y. Matsuo$^\ast$, R.-D. Zhu$^\ast$}
			\\[.4cm]
			{\em {}$^\dagger$Korea Institute for Advanced Study (KIAS)}\\
			{\em Quantum Universe Center (QUC)}\\
			{\em 85 Hoegiro, Dongdaemun-gu, Seoul, South Korea}
			\\[.4cm]
			{\em {}$^\ast$ Department of Physics, The University of Tokyo}\\
			{\em Bunkyo-ku, Tokyo, Japan}
			\\[.4cm]
		\end{center}
	\vfill
	\begin{abstract}
		Reflection states are introduced in the vertical and horizontal modules of the Ding-Iohara-Miki (DIM) algebra (quantum toroidal $\mathfrak{gl}_1$). Webs of DIM representations are in correspondence with $(p,q)$-web diagrams of type IIB string theory, under the identification of the algebraic intertwiner of Awata, Feigin and Shiraishi with the refined topological vertex. Extending the correspondence to the vertical reflection states, it is possible to engineer the $\CN=1$ quiver gauge theory of D-type (with unitary gauge groups). In this way, the Nekrasov instanton partition function is reproduced from the evaluation of expectation values of intertwiners. This computation leads to the identification of the vertical reflection state with the orientifold plane of string theory. We also provide a translation of this construction in the Iqbal-Kozcaz-Vafa refined topological vertex formalism. 
	\end{abstract}
	\vfill
\end{titlepage}

\section{Introduction}
The refined topological vertex, introduced independently by Iqbal, Kozcaz and Vafa (IKV) \cite{IKV} and Awata and Kanno \cite{Awata2008}, provides an efficient tool to compute the instanton partition function of $\CN=1$ (quiver) Super Yang-Mills (SYM) in the 5d background $\mathbb{R}^2_{\e_1}\times\mathbb{R}^2_{\e_2}\times S_R^1$.\footnote{Here $\mathbb{R}_\e^2$ denotes the Omega-deformation of $\mathbb{R}^2$ with equivariant parameter $\e$ \cite{Losev2003}.} The IKV vertex is a refinement with two parameters $(q,t)$ of the standard topological vertex used to compute amplitudes of topological strings on toric Calabi-Yau manifolds \cite{Leung-Vafa,aganagic2005topological,IKV}. The calculation of the instanton partition function is based on the $(p,q)$-brane web engineering of 5d SUSY gauge theories in type IIB string theory \cite{Hanany1996,AHK}. It can be performed using a Feynman-like diagrammatic technique, assigning a refined topological vertex $C_{\mu\nu\lambda}(q,t)$ to each junction, and summing over all the possible instanton configurations attached to the internal lines. The two parameters $q_1=t^{-1}$ and $q_2=q$ encode the dependence in the background parameters $\e_1,\e_2,R$.

The advent of the Alday-Gaiotto-Tachikawa (AGT) correspondence \cite{Alday2010,Wyllard2009} has brought a new perspective in the study of instanton partition functions of SYM theories. Indeed, the correspondence identifies the latter with the conformal blocks of a (q-deformed) W-algebra. It led to the study of the action of W-algebras on instanton partition functions, which has been expressed in terms of new quantum algebraic structures: the Spherical Hecke central algebra in the 4d case \cite{SHc}, and the Ding-Iohara-Miki (DIM) algebra in the 5d case \cite{DI,miki2007,KMZ}.\footnote{These two algebras can be seen alternatively as the affine Yangian of $\mathfrak{gl}_1$ \cite{Tsymbaliuk}, and the quantum toroidal $\mathfrak{gl}_1$ algebra respectively.} The presence of these algebraic structures has been instrumental for some of the proofs of the AGT correspondence, and further related the BPS sector of the gauge theories to quantum integrability \cite{Nekrasov2009,NPS}.

The deep relationship between the DIM algebra and the refined topological vertex has been elucidated by Awata, Feigin and Shiraishi (AFS) in \cite{AFS}. In this work, the topological vertex is identified with an intertwining operator between three modules of the DIM algebra.\footnote{For a usual affine algebras, such operators intertwine two Weyl modules and an evaluation representation. They play an important role in Wess-Zumino-Witten models and are known to obey the Knizhnik-Zamolodchikov equation \cite{Etingof:1998ru}.} More specifically, the DIM algebra possesses two types of representations relevant to SUSY gauge theories: horizontal (Fock) representations associated to NS5 branes, and vertical representations associated to D5 branes. The intertwiner defines a map between the three modules
\begin{equation}
\mbox{(Horizontal)$\ \otimes$ (Vertical)}\ \to\ \mbox{(Horizontal)}.
\end{equation} 
Matrix elements of the intertwiner coincide with the expression of the refined topological vertex in the appropriate basis. The three modules are associated to the three legs of the vertex, the vertical module corresponding to the preferred direction. In fact, the algebraic construction of AFS provides a free bosonic representation of the topological vertex.\footnote{In the unrefined case for which $\e_1+\e_2=0$, a fermionic construction had been previously proposed in \cite{Carlsson2013}.} With this method, the cumbersome summations over symmetric polynomials are replaced by simple commutations of oscillator modes.

The AFS construction of instanton partition functions has been further developed and generalized in \cite{Mironov:2016yue,Awata:2016riz,Awata2016,BFHMZ}. However, until now, this algebraic construction has been limited to A-type quiver gauge theories with unitary gauge groups. The main goal of this paper is to extend the description to quivers of D-type. For this purpose, we will introduce the notion of vertical and horizontal reflection states for DIM algebra, mimicking the boundary states of Virasoro algebra \cite{Ishibashi:1988kg}. These states are defined using two reflection symmetries of the DIM algebra related to the reflection of the $(p,q)$-web diagram along the (D5) and (NS5) directions. The vertical reflection state is then used to reproduce the instanton partition function of the D-type quiver.

A $(p,q)$-brane construction of D-type quiver SYM has been proposed in \cite{Kapustin-D,Hanany-Zaffaroni} by using the orientifold brane ${\bf ON}^0$. This construction can be further resolved using an ${\bf ON}^-$ brane 
\cite{HKLTY}. The corresponding brane-web picture leads to identify this orientifold brane with the vertical boundary state.

The paper is organized as follows. The second section is a brief reminder on the AFS construction of instanton partition functions using DIM algebra representation theory. We recall the definition of the algebra, its representations, and provide the expression of the intertwiner. In the third section, we define the reflection states in vertical and horizontal representation. Then, the calculation of the instanton partition function for the D-type quivers is presented in section four. Several relevant technical facts can be found in the appendix.

\vspace{1cm}

\textit{As this paper was in preparation, we became aware of a similar proposal by Sung-Soo Kim and Futoshi Yagi on the identification of the orientifold brane with a boundary state \cite{Kim2017}. Accordingly, we agreed to coordinate our release of the papers on arxiv.}

\section{Ding-Iohara-Miki Algebra and description of quiver gauge theories}
The correspondence between the representation theory of DIM algebra and the $(p,q)$-web construction of 5d $\CN=1$ quiver gauge theories \cite{Hanany1996,AHK} has been proposed in \cite{Mironov:2016yue,Awata:2016riz}. It has been generalized in \cite{BFHMZ} to include Chern-Simons couplings and stacks of D-branes. This section contains a brief reminder of this correspondence, using the conventions and notations of \cite{BFHMZ}.

\subsection{DIM algebra}
The algebra constructed by Ding-Iohara \cite{DI} and Miki \cite{miki2007} is the quantum toroidal deformation of $\mathfrak{gl}_1$. It can also be regarded as a one-parameter ($q,t$) deformation of the $W_{1+\infty}$-algebra. In the second Drinfeld presentation, the DIM algebra is engendered by the modes $\left\{x^\pm_k, \psi^{\pm}_l, \hg| k \in \mathbb{Z}, l \in \mathbb{Z}_{\geq0}\right\}$ with $\psi_0^+=(\psi_0^-)^{-1}$ and $\hg$ central elements. The algebra depends on the parameters $q_1, q_2, q_3 \in \mathbb{C}^\times$ constrained by relation $q_1q_2q_3 =1$. In the gauge theory, these parameters relate to the equivariant parameters of the Omega background $\e_1,\e_2$ and the radius $R$ of the compact dimension as follows:
\begin{equation}
q_1=e^{-R\e_1},\quad q_2=e^{-R\e_2}.
\end{equation} 
In the (refined) topological vertex formalism, these parameters are usually denoted $q_1 = t^{-1}, q_2 = q$.

The algebraic relations between the generators can be presented in terms of the Drinfeld currents 
\begin{equation}\label{Drinfelds}
x^\pm(z)=\sum_{k\in\mathbb{Z}}z^{-k}x^\pm_k,\quad \psi^+(z)=\sum_{k\geq0}z^{-k}\psi_k^+,\quad \psi^-(z)=\sum_{k\geq0}z^k\psi^-_{-k},
\end{equation}
they obey the following set of relations:
\begin{align}
\begin{split}\label{def_DIM}
&[\psi^\pm(z),\psi^\pm(w)]=0,\quad \psi^+(z)\psi^-(w)=\dfrac{g(\hg w/z)}{g(\hg^{-1}w/z)}\psi^-(w)\psi^+(z)\,,\\
&\psi^+(z)x^\pm(w)=g(\hg^{\mp1/2}w/z)^{\mp1}x^\pm(w)\psi^+(z),\quad \psi^-(z)x^\pm(w)=g(\hg^{\mp1/2}z/w)^{\pm1}x^\pm(w)\psi^-(z)\,,\\
&x^\pm(z)x^\pm(w)=g(z/w)^{\pm1}x^\pm(w)x^\pm(z)\,,\\
&[x^+(z),x^-(w)]=\dfrac{(1-q_1)(1-q_2)}{(1-q_1q_2)}\left(\d(\hg^{-1}z/w)\psi^+(\hg^{1/2}w)-\d(\hg z/w)\psi^-(\hg^{-1/2}w)\right).
\end{split}
\end{align}
The delta function here is defined as the formal sum $\delta(z)=\sum_{k\in\mathbb{Z}}z^k$, and the function $g(z)$ encodes the dependence in the parameters $q_1,q_2,q_3$ of the algebra:
\begin{equation}
g(z)=\prod_{a=1,2,3}\dfrac{1-q_a z}{1-q_a^{-1}z}.
\end{equation}

Like any affine algebra, the DIM algebra can be supplemented by grading operators. The algebra being twice affine, two grading operators can be introduced, denoted respectively $d$ and $\bd$:
\begin{align}
\begin{split}
&[d,x_k^\pm]=-kx_k^\pm,\quad [d,\psi_{\pm k}^\pm]=\mp k\psi_{\pm k}^\pm,\quad [d,\hg]=0,\\
&[\bd,x_k^\pm]=\pm x_k^\pm,\quad [\bd,\psi_{\pm k}^\pm]=0,\quad [\bd,\hg]=0,\quad [d,\bd]=0.
\end{split}
\end{align}
These grading operators can be employed to define two automorphisms of the DIM algebra, $\t_\a\cdot e= \a^d e\a^{-d}$ and $\bt_\a\cdot e= \a^{\bd} e\a^{-\bd}$ acting on any element $e$ of the algebra. Explicitly, these automorphisms act on the Drinfeld currents (\ref{Drinfelds}) as follows:
\begin{align}
\begin{split}
\t_\a\cdot x^\pm(z)=x^\pm(\a z),\quad \t_\a\cdot\psi^\pm(z)=\psi^\pm(\a z),\quad  \bt_\a\cdot x^\pm(z)=\a^{\pm1}x^\pm(z),\quad \bt_\a\cdot\psi^\pm(z)=\psi^\pm(z).
\end{split}
\end{align}
It is readily observed that the algebraic relations (\ref{def_DIM}) are indeed invariant under the action of these two automorphisms.

In addition to $\t_\a$ and $\bt_\a$, the DIM algebra is known to possess an $SL(2,\mathbb{Z})$ group of automorphisms generated by the elements $\CT$ and $\CS$ with $\CS^4=1=(\CS\CT)^3$. This group can be identified with the $SL(2,\mathbb{Z})$ group of duality in IIB string theory. The element $\CT$ acts on the generators as
\begin{equation}
\CT\cdot x^\pm_k=\hg^{\pm1}x_{k\mp1}^\pm,\quad \CT\cdot\psi^\pm_k=\hg^{\mp1}\psi^\pm_k.
\end{equation}
The second generator $\CS$ is called \textit{Miki's automorphism} \cite{miki2007}, its action takes a rather complicated form. However, the square of this automorphism has a closed action on the currents (\ref{Drinfelds}),
\begin{equation}
\CS^2\cdot x^\pm(z) = x^\mp(z^{-1}),\quad \CS^2\cdot \psi^\pm(z)=\psi^\mp(z^{-1}),\quad \CS^2\cdot\hg=\hg^{-1},
\end{equation} 
or, in terms of modes, $\CS^2\cdot x_k^\pm=x_{-k}^\mp$, and $\CS^2\cdot\psi_{\pm k}^\pm=\psi^\mp_{\mp k}$. The action of $\CS^2$ is involutive since $\CS^4=1$.

Finally, we would like to mention two involutive morphisms of algebra that sends the DIM algebra with parameters $q_1,q_2,q_3$ to a DIM algebra with inverted parameters $q_1^{-1},q_2^{-1},q_3^{-1}$ which will be referred to as $\overline{\mathrm{DIM}}$.  In effect, it induces a mapping $g(z)\rightarrow g(1/z)=g(z)^{-1}$. Corresponding to either choice $g(1/z)$ or $g(z)^{-1}$, two reflection automorphisms can be defined, denoted $\s_H$ and $\s_V$ respectively.\footnote{The reflection $\s_H$ has been previously denoted $\s_5$ in \cite{BFHMZ}.}  They are related by the action of Miki's automorphism as $\s_V=\CS^2\s_H$.Their action on DIM generators is given by
\begin{align}
\begin{split}
&\s_H\cdot x^\pm(z)=x^\pm(z^{-1}),\quad \s_H\cdot \psi^\pm(z)=\psi^\mp(z^{-1}),\quad \s_H\cdot\hg=\hg,\\
&\s_V\cdot x^\pm(z)=x^\mp(z),\quad \s_V\cdot \psi^\pm(z)=\psi^\pm(z),\quad \s_V\cdot\hg=\hg^{-1}.
\end{split}
\end{align}
These reflections will play an important role in the construction of the boundary states. 

\subsection{Representations}
The DIM algebra possesses two central elements that can be parametrized as
\begin{equation}
\hg=\g^{c},\quad \psi_0^+=(\psi_0^-)^{-1}=\g^{-\bc},
\end{equation} 
where we have introduced a shortcut notation for the parameter $\g=q_3^{1/2}$. In the application to $\CN=1$ SYM, we are interested in representations with integer central charges $(c,\bc)$, that will be denoted $\rho^{(c,\bc)}$. In particular, to a NS5-brane is associated a representation of central charges $(1,0)$, while a D5-brane is associated to a representation of central charges $(0,1)$. In general, a $(p,q)$-brane is seen as a stack of $p$ D5-branes and $q$ NS5-brane, to which corresponds a representation of central charges $(q,p)$. In the following, we will consider explicitly two different types of representations. Vertical representations have central charges $(0,m)$, they are associated to a stack of $m$ D5-branes related to a node of the quiver diagram with $U(m)$ gauge group. On the other hand, horizontal representations have central charges $(1,n)$, they are associated to a dyonic brane with one NS5 and $n$ D5-charge. The index $n$ is related to the Chern-Simons coupling of the gauge theory \cite{BFHMZ}.

From the action of the reflection symmetry $\s_H$ on the central element, we deduce that a representation $\rho^{(c,\bc)}$ of DIM transforms into a representation of central charges $(-c,\bc)$ of $\overline{\mathrm{DIM}}$. This leads to identify the reflection $\s_H$ as a reflection of the $(p,q)$-web diagram in the $(56)$-plane, in the direction leaving D5-branes invariant since the associated representations $(0,m)$ are invariant under $\s_H$. Similarly, $\s_V$ maps representations $(c,\bc)$ of DIM to representations $(c,-\bc)$ of $\overline{\mathrm{DIM}}$ leaving $(1,0)$ representations invariant. It is identified as the orthogonal reflection in the $(56)$-plane that leaves the NS5-brane direction invariant.

\paragraph{Notations} In order to differentiate the states of vertical and horizontal modules, we will use double ket states $|\cdots\rangle\rangle$ in the vertical case, and keep single ket states $|\cdots\rangle$ for the horizontal module.

\subsubsection{Vertical representations}
The modules of vertical representations $(0,m)$ possess a basis of states parametrized by a set of $m$ Young diagrams $\vec \lambda=(\lambda_1,\cdots,\lambda_m)$ \cite{FFJMM1,Feigin2011}. This basis will be denoted $\ket{\vec{v}, \vec\lambda}\rangle$, where $\vec v\in\mathbb{C}^{\otimes m}$ is the set of weights of the representation, i.e. the roots of the Drinfeld polynomial. The action of the DIM generators in this basis reads\footnote{These expressions can be found in \cite{BFHMZ}, up to an irrelevant modification of the states norm:
\begin{equation}
\aY\to \prod_{x\in\vec\l}(\chi_x/\g)^{-(m-1)}\aY.
\end{equation}}
\begin{align}\label{def_vert_rep}
\begin{split}
&\rho_{\vec v}^{(0,m)}(x^+(z))\ket{\vec v,\vec\lambda}\rangle =\sum_{x\in A(\vec\lambda)}\delta(z/\chi_x)\Res_{z=\chi_x}\dfrac1{z\cY_{\vec\lambda}(z)}|\vec v,\vec\lambda+x\rangle\rangle,\\
&\rho_{\vec v}^{(0,m)}(x^-(z))\ket{\vec 
v,\vec\lambda}\rangle=\gamma^{-m}\sum_{x\in R(\vec\lambda)}\delta(z/\chi_x)\Res_{z\rightarrow\chi_x}z^{-1}\cY_{\vec\lambda}(zq_3^{-1})\ket{\vec v,\vec \lambda-x}\rangle,\\
&\rho_{\vec v}^{(0,m)}(\psi^\pm(z))\ket{\vec v,\vec \lambda}\rangle=\gamma^{-m}\left[\Psi_{\lambda}(z)\right]_\pm\ket{\vec v,\vec\lambda}\rangle.
\end{split}
\end{align}
The actions of $x^\pm(z)$ involve a summation over the sets of boxes $A(\vec\l)$ (resp. $R(\vec\l)$) that can be added to $\vec\l$ (or removed from $\vec\l$). To each box $x=(l,i,j)$ in $\vec\l$ with coordinates $(i,j)$ in the $l$th Young diagram, has been associated the complex parameter $\chi_x=v_lq_1^{i-1}q_2^{j-1}\in\mathbb{C}$. Finally, the action has been written using the functions
\ba
\cY_{\vec\lambda}(z)=\frac{\prod_{x\in A(\vec\lambda)}1-\chi_x/z}{\prod_{x\in R(\vec\lambda)}1-\chi_xq_3^{-1}/z},\quad \Psi_{\vec\lambda}(z)=\frac{\cY_{\vec\lambda}(zq_3^{-1})}{\cY_{\vec\lambda}(z)},
\ea
and the subscript $\pm$ of the brackets $[\cdots]_\pm$ denotes an expansion in powers of $z^{\mp1}$.

The contragredient action $\hat\rho^{(0,m)}$ on the dual space with basis $\baY$ is defined as
\begin{equation}\label{def_hatrho}
\left(\baY\hat\rho^{(0,m)}(e)\right)|\vec v,\vec\l'\rangle\rangle=\baY\left(\rho^{(0,m)}(e)|\vec v,\vec\l'\rangle\rangle\right),
\end{equation}
for any element $e\in$DIM. Explicitly, the action on the dual states read as in (\ref{def_vert_rep}) with $x^+$ and $x^-$ exchanged (and the sign flipped).\footnote{Again, the normalization of the dual states have been modified with respect to \cite{BFHMZ},
\begin{equation}
\baY\to \g^{-(m-1)|\vec\l|}\baY.
\end{equation}} In agreement with the representation, the states are normalized as
\begin{equation}\label{def_a}
\baY\vec v,\vec\l'\rangle\rangle=a_{\vec\l}^{-1}\d_{\vec\l,\vec\l'},\quad a_{\vec\l}=\Zv(\vec v,\vec\l)\prod_{l=1}^m(-\g v_l)^{-|\vec\l|}\prod_{x\in\vec\l}\chi_x^m,
\end{equation}
where $\Zv$ is the vector multiplet contribution to the instanton partition function (its expression can be found in Appendix \ref{secB}). Note that the coefficients $a_{\vec\l}$ are invariant under the rescaling of the weights $\vec v\to\a\vec v$.

In the vertical action (\ref{def_vert_rep}), the dependence on the spectral parameter $z$ is written in terms of the ratios $z/\chi_x$. As a result, any rescaling of the spectral parameter can be compensated by a similar rescaling of the weights of the gauge theory. Thus, the action of the automorphism $\t_\a$ can be rephrased as a collective rescaling of the weights, $\rho^{(0,m)}_{\vec v}(\t_\a\cdot e)=\rho^{(0,m)}_{\a^{-1}\vec v}(e)$ for $e\in$DIM. On the other hand, the action of the automorphism $\bt_\a$ is equivalent to a dilatation, $\aY\to \a^{-|\vec\l|}\aY$.

\subsubsection{Horizontal representation}
The horizontal representation $(1,n)$ \cite{Feigin2009a} acts on the Fock module of a q-deformed bosonic field with modes\footnote{These modes have been rescaled as $a_k\to (1-t^k)\g^{|k|/2}a_k$ with respect to the conventions used in \cite{BFHMZ}, and we have set $q_1=t^{-1}$, $q_2=q$. In this new form, the commutation relation (\ref{q-osc}) is invariant under the inversion of the parameters $q_a\to q_a^{-1}$.}
\begin{equation}\label{q-osc}
[a_k,a_l]=c_k\d_{k+l},\quad\text{with}\quad c_k=k(1-q_1^{|k|})(1-q_2^{|k|})q_3^{|k|/2}.
\end{equation}
The vacuum state $|\vac\rangle$ is annihilated by the positive modes $a_{k>0}$. We also define the normal ordering by placing the positive modes on the right of the negative ones. Then, the horizontal action of DIM algebra with weight $u$ can be expressed using the vertex operators
\begin{align}
\begin{split}\label{1nrep}
&\rho^{(1,n)}_u(x^\pm(z))=u^{\pm1} \g^{\pm n}z^{\mp n}\eta^\pm(z),\quad \rho^{(1,n)}_u(\psi^\pm(z))=\g^{\mp n}\vphi^\pm(z),\quad \rho^{(1,n)}_u(\hg)=\g\\
&\text{with}\quad \eta^\pm(z)=:\exp\left(\mp\sum_{k\neq0}\dfrac{z^{-k}}{k}\g^{\mp|k|/2}a_k\right):,\quad\vphi^\pm(z)=\exp\left(\pm\sum_{k>0}\dfrac{z^{\mp k}}{k}(\g^k-\g^{-k})a_{\pm k}\right).
\end{split}
\end{align}

The contragredient action is given as usual by $(a_k)^\dagger=a_{-k}$. Combined with the sign flip of the modes, i.e. $a_k\to -a_{-k}$, it realizes the representation of the horizontal reflection $\s_H$ of the algebra, as $\eta^\pm(z)\to \eta^\pm(z^{-1})$ and $\vphi^\pm(z)\to\vphi^\mp(z^{-1})$. On the other hand, the vertical reflection $\s_V$ takes the simple form of a sign flip of the modes $a_k\to-a_k$.

Finally, we examine the action of the automorphism $\t_\a$ and $\bt_\a$ in the horizontal representation. The action of $\t_\a$ corresponds to a rescaling of the modes $a_k\to\a^{-k} a_k$ which leaves the commutation relation (\ref{def_a}) invariant. On the other hand, $\bt_\a$ acts as a rescaling of the weights $u\to \a u$, i.e. $\rho_u^{(1,n)}(\t_\a\cdot e)=\rho_{\a u}^{(1,n)}(e)$.

\subsection{Generalized Awata-Feigin-Shiraishi intertwiner}\label{rel_to_tv}

Intertwiners between horizontal and vertical representations of DIM algebra have been defined by Awata, Feigin and Shiraishi in \cite{AFS}. This definition has been generalized to higher ranks of vertical representations in \cite{BFHMZ}. The intertwiner $\Phi^{(n,m)}[u,\vec v]$ maps the tensor product of an horizontal and a vertical module to states in a horizontal module. It is obtained as the unique solution (up to normalization) of the equation
\begin{align}
\begin{split}
&\Phi^{(n,m)}[u,\vec v]:(0,m)_{\vec v}\otimes(1,n)_u\to(1,n+m)_{u'},\\
&\rho_{u'}^{(1,n+m)}(e)\Phi^{(n,m)}[u,\vec v]=\Phi^{(n,m)}[u,\vec v]\cdot\left(\rho_{\vec v}^{(0,m)}\otimes\rho_u^{(1,n)}\right)\D(e),
\end{split}
\end{align}
where $\D$ is the coproduct given in (\ref{AFS_coproduct}) and the weight of the horizontal representation in the target space is fixed to $u'=u\prod_{l=1}^m(-v_l)$. The intertwiner can be seen as a vector in the vertical module, with components given by operators acting in the Fock modules. Their explicit expression has been worked out in \cite{AFS,BFHMZ} and read
\begin{align}
\begin{split}
&\Phi^{(n,m)}_{\vec\l}[u,\vec v]=\Phi^{(n,m)}[u,\vec v]\aY,\\
&\Phi^{(n,m)}_{\vec\l}[u,\vec v]=:\prod_{l=1}^m\Phi_\vac(v_l)\prod_{x\in\vec\l}\rho_{u'}^{(1,n+m)}(x^+(\chi_x)):,\quad \Phi_\vac(v)=:\exp\left(\sum_{k\neq0}\dfrac{(\g v)^{-k}}{c_k}\g^{-|k|/2}a_k\right):
\end{split}
\end{align}
It can be written more explicitly in terms of the vertex operator $\eta^+(z)$,
\begin{equation}\label{def_t}
\Phi^{(n,m)}_{\vec\l}[u,\vec v]=t_{n,m}(\vec\l,u,\vec v):\prod_{l=1}^m\Phi_\vac(v_l)\prod_{x\in\vec\l}\eta^+(\chi_x):,\quad t_{n,m}(\vec\l,u,\vec v)=(u')^{|\vec\l|}\prod_{x\in\vec \l}(\chi_x/\g)^{-m-n},
\end{equation} 
where we have singled out the normalization factor $t_{n,m}(\vec\l,u,\vec v)$.

The dual intertwiner is defined in a similar way as the unique solution to the problem,
\begin{align}
\begin{split}
&\Phi^{(n,m)\ast}[u,\vec v]:(1,n+m)_{u'}\to(1,n)_u\otimes(0,m)_{\vec v},\\
&\Phi^{(n,m)\ast}[u,\vec v]\rho_{u'}^{(1,n+m)}(e)=\left(\rho_u^{(1,n)}\otimes\rho_{\vec v}^{(0,m)}\right)\D(e)\cdot\Phi^{(n,m)\ast}[u,\vec v].
\end{split}
\end{align}
The vertical projections are operators in the horizontal modules:
\begin{align}
\begin{split}
\Phi^{(n,m)\ast}_{\vec\l}[u,\vec v]&=\baY\Phi^{(n,m)\ast}[u,\vec v],\\
\Phi^{(n,m)\ast}_{\vec\l}[u,\vec v]&=\g^{-m|\vec\l|}:\prod_{l=1}^m\Phi_\vac^\ast(v_l)\prod_{x\in\vec\l}\rho_{u}^{(1,n)}(x^-(\chi_x)):,\quad \Phi_\vac^\ast(v)=:\exp\left(-\sum_{k\neq0}\dfrac{(\g v)^{-k}}{c_k}\g^{|k|/2}a_k\right):\\
&=t_{n,m}^\ast(\vec\l,u,\vec v):\prod_{l=1}^m\Phi_\vac^\ast(v_l)\prod_{x\in\vec\l}\eta^-(\chi_x):,\quad t_{n,m}^\ast(\vec\l,u,\vec v)=(\g^m u)^{-|\vec\l|}\prod_{x\in\vec\l}(\chi_x/\g)^n.
\end{split}
\end{align}

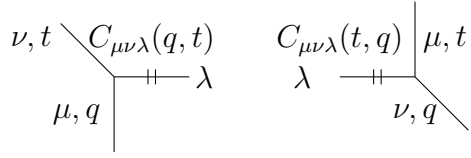
\begin{figure}
\begin{center}
\begin{tikzpicture}
		\begin{scope}[xscale = -1]
\draw (-0.71,-0.71)-- (0,0);
\draw (1,0)--(0,0);
\draw (0,0)--(0,1);
\node at (1,0.5) {$C_{\mu\nu\lambda}(t,q)$};
\node at (-0.4,0.5) {$\mu,t$};
\node at (0,-0.5) {$\nu,q$};
\node at (1.5,0) {$\lambda$};
\draw (0.45,0.1)--(0.45,-0.1);
\draw (0.55,0.1)--(0.55,-0.1);
\draw (3,0)-- (4,0);
\draw (4,0)--(4,-1);
\draw (4,0)--(4.71,0.71);
\node at (3.5,0.5) {$C_{\mu\nu\lambda}(q,t)$};
\draw (3.45,0.1)--(3.45,-0.1);
\draw (3.55,0.1)--(3.55,-0.1);
\node at (4.5,-0.5) {$\mu,q$};
\node at (5.1,0.5) {$\nu,t$};
\node at (2.8,0) {$\lambda$};
\end{scope}
\end{tikzpicture}
\end{center}
\caption{Representation of the topological vertex.\label{f:IKV}}
\end{figure}

\begin{figure}
	\begin{center}
		\begin{tikzpicture}
		\begin{scope}[xscale = -1]
		\node[below,scale=0.7] at (1,-1) {$(1,n)_u$};
		\node[right,scale=0.7] at (0,0) {$(0,m)_{\vec v}$};
		\node[above,scale=0.7] at (1,1) {$(1,n+m)_{u'}$};
		\node[left,scale=0.7] at (1,0) {$\Phi^{(n,m)}[u,\vec v]$};
		\draw (0,0) -- (1,0) -- (1,1);
		\draw (1,0) -- (1,-1);
		\end{scope}
		\begin{scope}[shift={(4,1)},xscale = -1]
		\node[below,scale=0.7] at (0,-2) {$(1,n+m)_{u'}$};
		\node[above,scale=0.7] at (0,0) {$(1,n)_u$};
		\node[left,scale=0.7] at (1,-1) {$(0,m)_{\vec v}$};
		\node[right,scale=0.7] at (0,-1) {$\Phi^{(n,m)\ast}[u,\vec v]$};
		\draw (0,0) -- (0,-1) -- (1,-1);
		\draw (0,-1) -- (0,-2);
		\end{scope}
		\begin{scope}[shift={(8,-1)},scale=0.5]
		\draw[->] (0,0)--(0,1);
		\draw[->] (0,0)--(1,0);
		\node[scale=0.7,above] at (0,1) {horizontal};
		\node[scale=0.7,right] at (1,0) {vertical};
		\end{scope}
		\end{tikzpicture}
		
	\end{center}
	\caption{$\Phi^{(n,m)}[u,\vec v]$ and $\Phi^{(n,m)\ast}[u,\vec v]$}
	\label{fig_vertex}
\end{figure}
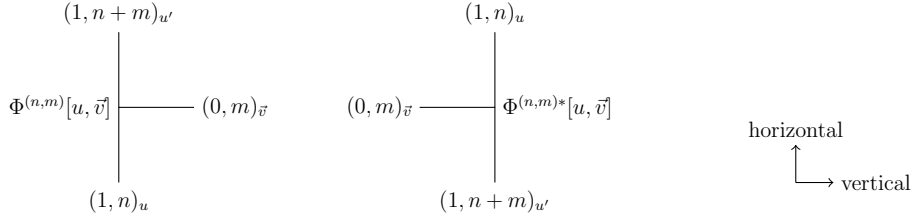

\begin{figure}
\begin{center}
\begin{tikzpicture}[scale=0.5]
\draw (0,0)--(1,1)--(1,3)--(0,4);
\draw (4,0)--(3,1)--(3,3)--(4,4);
\draw (1,1)--(3,1);
\draw (1,3)--(3,3);
\draw[ultra thick,->] (7.5,2)--(9,2);
\begin{scope}[shift={(5.5,0)}]
\draw[->] (0,0)--(0,1);
\draw[->] (0,0)--(1,0);
\node[scale=0.5,above] at (0,1) {$6$};
\node[scale=0.5,right] at (1,0) {$5$};
\end{scope}
\begin{scope}[shift = {(11,0.4)}, scale = 0.8]
\draw(0,0)--(0,4);
\draw(0,2)--(4,2);
\draw(4,0)--(4,4);
\node[scale=0.7,below] at (0,0) {$(1,n)$};
\node[scale=0.7,below] at (4,0) {$(1,n^\ast+2)$};
\node[scale=0.7,above] at (0,4) {$(1,n+2)$};
\node[scale=0.7,above] at (4,4) {$(1,n^\ast)$};
\node[scale=0.7,above] at (2,2) {$(0,2)$};
\node[scale=0.7,left] at (0,2) {$\Phi$};
\node[scale=0.7,right] at (4,2) {$\Phi^\ast$};

\begin{scope}[scale=1.25,shift={(0.5,-0.7)}]
\draw[->] (6,0)--(6,1);
\draw[->] (6,0)--(7,0);
\node[scale=0.5,above] at (6,1) {horizontal};
\node[scale=0.5,right] at (7,0) {vertical};
\end{scope}
\end{scope}
\end{tikzpicture}
\hspace{1cm}

\end{center}
\caption{$(p,q)$-brane web diagram of $A_1$ $U(2)$ quiver gauge theory (left) and the corresponding web of DIM representations (right)}
\label{fig_U2}
\end{figure}
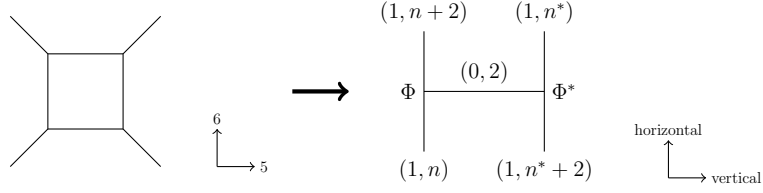

The original AFS intertwiner is recovered by restriction to a vertical representation of rank $m=1$. This intertwiner is then identified with the refined topological vertex \cite{IKV,Awata2008}, the vertical module being associated to the preferred direction \cite{AFS}. The topological vertex is represented in Figure \ref{f:IKV}, and a brief reminder is given in Appendix \ref{secC}. By construction, the topological vertex describes the intersection of a D5 brane, an NS5 brane and a dyonic $(1,1)$ brane.  The generalized intertwiners describe the intersection of a stack of $m$ D5 branes with a dyonic brane with one NS5 and $n$ D5 charge.  They are represented graphically in Figure \ref{fig_vertex}. 
The figure of brane web represents intersections of $(p,q)$-branes in the $(56)$-plane, assuming that D5 branes extend in the spacetime directions $012345$ while NS5 branes extend in the directions $012346$. For the representation web of DIM using generalized AFS vertices, the horizontal line may represent more than one D-brane. To avoid complication, we use vertical (resp. horizontal) lines to describe horizontal (resp. vertical) representations (see Figure \ref{fig_U2}). 

The generalized intertwiner of rank $m$ is obtained by taking the product of $m$ AFS intertwiners (or $m$ topological vertices) along the horizontal (i.e. NS5 or non-preferred) directions. Following the rules (\ref{contraction_rule}) given below, the normal ordering produces a product of Nekrasov factors that coincides through the formula (\ref{Zbf}) with the vector multiplet contribution $\Zv$ to the instanton partition function. This contribution has been absorbed in the normalization of the vertical states, as can be seen in the definition (\ref{def_a}) of the factor $a_{\vec\l}$.  The main advantage of working with generalized intertwiners is to be able to replace the tensor product of $m$ $(0,1)$ vertical representations by a single $(0,m)$ representation, thus enabling an easier treatment of gauge groups $U(m)$ with arbitrary rank $m$. Along these lines, we will replace the $(p,q)$-brane web diagram of a quiver gauge theory by a simpler diagram, the \text{DIM representation web diagram}. In this simplified diagram, the product of topological vertices is replaced by a generalized intertwiner, thus representing a stack of D5 branes by a single edge. In addition, any oblique line corresponding to a $(1,n)$ representation will be drawn vertically. For instance, the $(p,q)$-brane web diagram of a $U(2)$ $A_1$ gauge theory (Figure \ref{fig_U2} left) is replaced by the diagram on the right of Figure \ref{fig_U2}. This diagram takes the same form for any gauge group $U(m)$, at the condition of replacing the representations $(0,2)$, $(1,n+2)$,... with their rank $m$ counterparts.

\subsection{Webs of DIM representations}\label{sec_web}
Instanton partition functions of quiver $\CN=1$ 5d SYM can be computed from the $(p,q)$-brane web diagram by assigning a topological vertex to the brane junctions, and applying the appropriate gluing rules \cite{IKV,Taki07}. The AFS intertwiners provide a free bosonic representation of the topological vertex, and similar gluing rules apply to them. More precisely, topological vertices can be coupled in two different directions. The coupling in the D5 channel, or preferred direction, corresponds to a scalar product of the intertwiners in the vertical space. It can be computed by insertion of the closure relation of the vertical basis,
\begin{align}
\begin{split}
\Phi^{(n,m)}[u,\vec v]\cdot\Phi^{(n^\ast,m)\ast}[u^\ast,\vec v]&=\sum_{\vec\l}a_{\vec\l}\ \Phi^{(n,m)}[u,\vec v]\aY\baY\Phi^{(n^\ast,m)\ast}[u^\ast,\vec v]\\
&=\sum_{\vec\l}a_{\vec\l}\ \Phi^{(n,m)}_{\vec\l}[u,\vec v]\otimes\Phi^{(n^\ast,m)\ast}_{\vec\l}[u^\ast,\vec v].
\end{split}
\end{align}
The vacuum expectation value of this operator, obtained by considering the vacuum state component in each horizontal module, reproduces the instanton partition function of a pure $U(m)$ gauge theory with Chern-Simons coupling $\k=n^\ast-n$,
\begin{align}
\begin{split}
\la\Phi^{(n,m)}[u,\vec v]\cdot\Phi^{(n^\ast,m)\ast}[u^\ast,\vec v]\ra&=\left(\langle\vac|\otimes\langle\vac|\right)\Phi^{(n,m)}[u,\vec v]\cdot\Phi^{(n^\ast,m)\ast}[u^\ast,\vec v]\left(|\vac\rangle\otimes|\vac\rangle\right)\\
&=\sum_{\vec\l}a_{\vec\l}\ \langle\vac|\Phi^{(n,m)}_{\vec\l}[u,\vec v]|\vac\rangle\langle\vac|\Phi^{(n^\ast,m)\ast}_{\vec\l}[u^\ast,\vec v]|\vac\rangle\\
&=\sum_{\vec\l}a_{\vec\l}\ t_{n,m}(\vec\l,u,\vec v) t_{n^\ast,m}^\ast(\vec\l,u^\ast,\vec v)\\
&=\sum_{\vec\l} \qf^{|\vec\l|}\Zv(\vec v,\vec\l)\ZCS(\k,\vec\l).
\end{split}
\end{align}
This computation leads to identify the weights with the position of the branes. Indeed, the weights in the vertical representation, $\vec{v}$, coincide with the (exponentiated) Coulomb branch vevs. On the other hand, the exponentiated gauge coupling coincides with a ratio of the horizontal weights, $\qf=\g^{-\k-m}u/u^\ast$.

The second possibility is to couple the intertwiners along the NS5-direction, or in fact any oblique direction corresponding to a bound state of the NS5 brane with several D5 branes. This type of gluing is realized as a product in the Fock modules of the horizontal representations. The normal-ordering of this product generates the instanton bifundamental contributions necessary to build quiver gauge theories:\footnote{In the correspondence with the IKV topological vertex, these contractions correspond to the summation rule of the Schur polynomial, 
\ba
\sum_\lambda s_{\lambda^t}(x)s_\lambda(y)=\prod_{i,j}(1+x_iy_j).
\ea}
\begin{align}
\begin{split}\label{contraction_rule}
&\Phi_{\vec\l_1}^{(n,m_1)}[u,\vec v_1]\Phi_{\vec\l_2}^{(n,m_2)\ast}[u,\vec v_2]=\bZbf(\vec v_2,\vec \l_2,\vec v_1,\vec \l_1|\g^{-1}):\Phi_{\vec\l_2}^{(n,m_2)\ast}[u,\vec v_2]\Phi_{\vec\l_1}^{(n,m_1)}[u,\vec v_1]:,\\
&\Phi_{\vec\l_2}^{(n_2,m_2)\ast}[u_2,\vec v_2]\Phi_{\vec\l_1}^{(n_1,m_1)}[u_1,\vec v_1]=\bZbf(\vec v_1,\vec \l_1,\vec v_2,\vec \l_2|\g^{-1}):\Phi_{\vec\l_2}^{(n_2,m_2)\ast}[u_2,\vec v_2]\Phi_{\vec\l_1}^{(n_1,m_1)}[u_1,\vec v_1]:,\\
&\Phi_{\vec\lambda_2}^{(n_1+m_1,m_2)}[u_1',\vec v_2]\Phi_{\vec\lambda_1}^{(n_1,m_1)}[u_1,\vec v_1]=\bZbf(\vec v_1,\vec\lambda_1,\vec v_2,\vec\lambda_2|1)^{-1} :\Phi_{\vec\lambda_2}^{(n_1+m_1,m_2)}[u_1',\vec v_2]\Phi_{\vec\lambda_1}^{(n_1,m_1)}[u_1,\vec v_1]:\,,\\
&\Phi_{\vec\l_2}^{(n_1-m_2,m_2)\ast}[u_2,\vec v_2]\Phi_{\vec\l_1}^{(n_1,m_1)\ast}[u_1,\vec v_1]=\bZbf(\vec v_1,\vec\lambda_1,\vec v_2,\vec\lambda_2|\g^{-2})^{-1} :\Phi_{\vec\l_2}^{(n_1-m_2,m_2)\ast}[u_2,\vec v_2]\Phi_{\vec\l_1}^{(n_1,m_1)\ast}[u_1,\vec v_1]:.
\end{split}
\end{align}
To lighten the notations, we have introduced the rescaled quantity
\begin{equation}
\bZbf(\vec v_1,\vec\l_1,\vec v_2,\vec\l_2|\mu)=\dfrac{\Zbf(\vec v_1,\vec\l_1,\vec v_2,\vec\l_2|\mu)}{\prod_{l_1=1}^{m_1}\prod_{l_2=1}^{m_2}\CG(v_{l_1}^{(1)}/(q_3\mu v_{l_2}^{(2)}))}.
\end{equation} 
The function $\CG(z)$ is independent of the instanton configurations, it is interpreted as a perturbative factor that can be disregarded in our discussion of the instanton contributions (its expression can be found in \cite{BFHMZ}). Using these contractions,  the instanton partition function of linear quivers is reproduced by the (Fock) vacuum expectation value of the product of intertwiners following the prescription of the $(p,q)$-web diagram \cite{BFHMZ}.

\section{Reflection states of the Ding-Iohara-Miki algebra}
Our proposal for the reflection states in DIM algebra is inspired by the boundary states of Virasoro algebra \cite{Ishibashi:1988kg} that play an important role in 2d boundary conformal field theories \cite{Cardy:1989ir}. These states live in the tensor product of two Verma modules, and satisfy\footnote{This condition is not restrictive enough to define these states, and one should rely on the Sugawara construction and impose the condition $(J_n^a\otimes1+1\otimes J_{-n}^a)|B\rangle =0$ \cite{Ishibashi:1988kg}.}
\ba\label{bound-def}
(L_n\otimes1-1\otimes L_{-n})|\Omega\rangle =0.
\ea
In the second term, $L_{-n}=(L_n)^\dagger$ corresponds to the contragredient action of Virasoro generators. Note the presence of the morphism $\s\cdot L_n=L_{-n}$ that sends the Virasoro algebra with central charge $c$ to a Virasoro algebra with central charge $-c$. 

Compared to the Virasoro algebra, the DIM algebra has a richer (auto)morphism structure, and we expect a large class of boundary states. In this paper, we will not attempt to a general classification of boundary states but present only the simplest constructions associated with the horizontal and vertical representations. We hope to come back to the classification issue in the near future. 

The action of DIM algebra on instanton partition functions is very different from the action of Virasoro algebra on 2d conformal field theory. In our context, the interpretation of the states satisfying (\ref{bound-def}) as boundary states seems misleading, and we will use instead the terminology \textit{reflection state}.

\subsection{Vertical reflection state}
By analogy with Virasoro boundary states, we are looking for states in the tensor product of two vertical modules and satisfying the constraint
\begin{equation}\label{prop_vert_st}
\left(\rho^{(0,m)}(e)\otimes1\right)|\Omega\rangle\rangle=\left(1\otimes\bar\rho^{(0,m)}(e)\right)|\Omega\rangle\rangle,\quad e\in\text{DIM}.
\end{equation} 
The RHS involves the dual action that has been expressed in terms of the reflection symmetry $\s_V$.
\begin{equation}\label{adj_vert_rep}
\bar\rho^{(0,m)}(e)=(-)^{\e(e)-1}\rho^{(0,m)}(\s_V\cdot e),
\end{equation} 
where $\e$ is the co-unit, $\e(x^\pm_k)=0$, $\e(\psi^\pm_{\pm k})=1$.  $\bar{\rho}$ provides a representation of $\overline{\mathrm{DIM}}$, it is the transposed of the contragredient representation $\hat\rho$ defined in (\ref{def_hatrho}). More explicitly, the characterization (\ref{prop_vert_st}) of the boundary state in terms of the action of the Drinfeld currents (\ref{Drinfelds}) reads
\begin{align}\label{rs-prop}
\begin{split}
&(\rho^{(0,m)}(x^{+}(z))\otimes 1)|\Omega\rrangle\rangle = -(1\otimes \rho^{(0,m)}(x^{-}(z)))|\Omega\rrangle\rangle\,,\\
&(\rho^{(0,m)}(x^{-}(z))\otimes 1)|\Omega\rrangle\rangle = -(1\otimes  \rho^{(0,m)}(x^{+}(z)))|\Omega\rrangle\rangle\,,\\
&(\rho^{(0,m)}(\psi^{\pm}(z))\otimes 1)|\Omega\rrangle\rangle = (1\otimes \rho^{(0,m)}(\psi^{\pm}(z)))|\Omega\rrangle\rangle\,.
\end{split}
\end{align}
These constraints are satisfied by a coherent state given by
\begin{equation}\label{def_Omega}
|\Omega\rangle\rangle=\sum_{\vec \lambda}a_{\vec \lambda}\ \aY\otimes\aY.
\end{equation}
Indeed, this state is obtained from the identity operator 
\begin{equation}
{\rm id}=\sum_{\vec\lambda} a_{{\vec\lambda}} \ket{\vec{v},\vec{\lambda}}\rangle\langle\bra{\vec{v},\vec{\lambda}},
\end{equation} 
with the second state has been transposed. The identities (\ref{prop_vert_st}) are the equivalent of $\rho^{(0,m)}(e) \cdot \mbox{id}= \mbox{id}\cdot \hat\rho^{(0,m)}(e)$ with $\hat\rho$ replaced by its transpose $\hat\rho^t=\bar\rho$.

In the definition (\ref{def_Omega}) of the boundary state, we have assumed that the two vertical modules have the same weights. In the following, we will need a more general definition that relaxes this condition,
\begin{equation}\label{ver_ref_ket}
|\Omega,\a\rangle\rangle=\sum_{\vec \lambda}a_{\vec \lambda}\ \aY\otimes|\a\vec v,\vec\l\rangle\rangle.
\end{equation} 
From the scaling property of the vertical representation with respect to the automorphism $\t_\a$, it can be shown that this state satisfies the property
\begin{equation}
\left(\rho^{(0,m)}(e)\otimes1\right)|\Omega,\a\rangle\rangle=\left(1\otimes\bar\rho^{(0,m)}(\t_\a\cdot e)\right)|\Omega,\a\rangle\rangle,\quad e\in\text{DIM}.
\end{equation} 

\begin{figure}
\begin{center}
	\begin{tikzpicture}
	\draw (0,0.3)--(-1,0.3);
	\draw (0,0)--(-1,0);
	\draw[dashed] (0,1)--(0,-1);
	\node at (3,0) {$\leftrightarrow\qquad|\Omega,\a\rangle\rangle$};
	\end{tikzpicture}
\end{center}
\caption{Graphical representation of the boundary state in the $(p,q)$-web diagram.\label{fig_bdy_state}}
\end{figure}
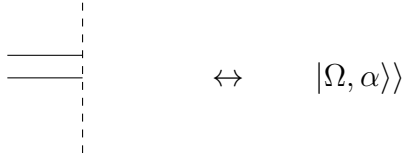

In the correspondence with $(p,q)$-web diagrams, the boundary state will be associated to an orientifold plane that realizes the vertical reflection $\s_V$. Graphically, it will be represented as a dashed line in the brane web diagram (see Figure \ref{fig_bdy_state}).

We would like to conclude this section with two important remarks. Firstly, the vertical representation of rank $m=1$ can be rephrased as an action of the DIM algebra on Macdonald symmetric polynomials under the identification $|v,\l\rangle\rangle\sim P_\l(x)$ \cite{FFJMM1}. In this context, the reflection state has the natural interpretation of the Macdonald kernel function,
\begin{eqnarray}
|\Omega\rangle\rangle \sim \sum_{\lambda}\dfrac{P_\lambda(x)P_\lambda(y)}{\langle P_\lambda,P_\lambda\rangle}=\prod_{i,j}\frac{(tx_i y_j;q)_\infty}{(x_i y_j;q)_\infty}. 
\end{eqnarray}

Secondly, the vertical representation of DIM algebra contains an action of a q-deformed $\cW_m$ algebra coupled to a Heisenberg algebra (the so-called $U(1)$-factor). For $m=2$, the reflection state (\ref{def_Omega}) defines a boundary state for the q-Virasoro algebra \cite{Shiraishi1995}. In the degenerate limit $q_2=q_1^{-\b}\to1$, the DIM algebra reduces to the Spherical Hecke central algebra \cite{SHc} which share a similar relation with $\cW_m$ algebras. In particular, for $m=2$, the vertical reflection state satisfies the condition of a Virasoro boundary state.\footnote{To be more specific, the degenerate version of the state (\ref{def_Omega}) satisfies $(L_n\otimes1)|\Omega\rangle\rangle=(-1)^n(1\otimes L_{-n})|\Omega\rangle\rangle$. In order to reproduce (\ref{bound-def}) an extra twist of $\bt_\a$ is needed, effectively replacing $a_{\vec\l}$ with $(-1)^{|\vec\l|}a_{\vec\l}$ in (\ref{def_Omega}).}

\subsection{Horizontal reflection state}
Both the horizontal representation of DIM and the $c=1$ representations of Virasoro algebra act on the Fock module of a free bosonic field. In the latter case, the boundary state has been worked out in \cite{Callan:1987px,Polchinski:1987tu}. Taking into account the q-deformation, we note that the state
\begin{equation}
|\Omega\rangle=\exp\left(-\sum_{k>0}\dfrac1{c_k}\ a_{-k}\otimes a_{-k}\right)|\vac\rangle\otimes|\vac\rangle,
\end{equation}
where the coefficients $c_k$ are taken from the commutation relations (\ref{def_a}), satisfies the reflection condition 
\begin{equation}
\left(a_k\otimes 1+1\otimes a_{-k}\right)|\Omega\rangle=0.
\end{equation} 
As a consequence of this identity, the action of the vertex operators $\eta^\pm(z)$ and $\vphi^\pm(z)$ obey
\begin{equation}
\left(\eta^\pm(z)\otimes1\right)|\Omega\rangle=\left(1\otimes\eta^\pm(z^{-1})\right)|\Omega\rangle,\quad\left(\vphi^\pm(z)\otimes1\right)|\Omega\rangle=\left(1\otimes\vphi^\mp(z^{-1})\right)|\Omega\rangle,
\end{equation} 
Thus, in the horizontal representation $(1,0)$, this boundary state is associated to the horizontal reflection $\s_H$:
\begin{equation}\label{hor_state}
\left(\rho_u^{(1,0)}(e)\otimes1\right)|\Omega\rangle=\left(1\otimes\rho_u^{(1,0)}(\s_H\cdot e)\right)|\Omega\rangle,\quad e\in\text{DIM}.
\end{equation} 

Just like the vertical reflection state, the horizontal reflection state can also be twisted by the automorphisms $\t_\a$ and $\bt_\a$. The twist by $\bt_a$ is obtained by choosing the weight $\a u$ instead of $u$ for the horizontal representation in the second Fock module. However, the twist by the orthogonal grading $\t_\a$ seems more interesting. It is obtained by rescaling the oscillator modes in the second module, defining
\begin{equation}
|\Omega,\a\rangle=\exp\left(-\sum_{k>0}\dfrac{\a^k}{c_k}\ a_{-k}\otimes a_{-k}\right)|\vac\rangle\otimes|\vac\rangle\implies \left(a_k\otimes 1+\a^k\ 1\otimes a_{-k}\right)|\Omega,\a\rangle=0.
\end{equation} 
Consequently, the identity (\ref{hor_state}) is modified into
\begin{equation}
\left(\rho_u^{(1,0)}(e)\otimes1\right)|\Omega,\a\rangle=\left(1\otimes\rho_u^{(1,0)}(\t_\a \cdot\s_H\cdot e)\right)|\Omega,\a\rangle,\quad e\in\text{DIM}.
\end{equation}

\section{$D$-type Quiver and vertical reflection}
\begin{figure}
\begin{center}
\begin{tikzpicture}[scale=0.4]
\draw (0,0) circle (1);
\node[scale=0.7] at (0,0) {$1$};
\draw (5,0) circle (1);
\node[scale=0.7] at (5,0) {$2$};
\draw (10,0) circle (1);
\node[scale=0.7] at (10,0) {$3$};
\draw (16,0) circle (1);
\node[scale=0.7] at (16,0) {$r-2$};
\draw (19.8,2.8) circle (1);
\node[scale=0.7] at (19.8,2.8) {$r-1$};
\draw (19.8,-2.8) circle (1);
\node[scale=0.7] at (19.8,-2.8) {$r$};
\draw[<-] (1,0)--(4,0);
\draw[<-] (6,0)--(9,0);
\draw[<-] (11,0)--(12,0);
\draw[dotted] (12,0)--(14,0);
\draw (14,0)--(15,0);
\draw[<-] (17,0)--(19.1,2.1);
\draw[->] (17,0)--(19.1,-2.1);
\end{tikzpicture}
\caption{Dynkin diagram of a $D_r$ Lie algebra}
\label{fig_Dr}
\end{center}
\end{figure}
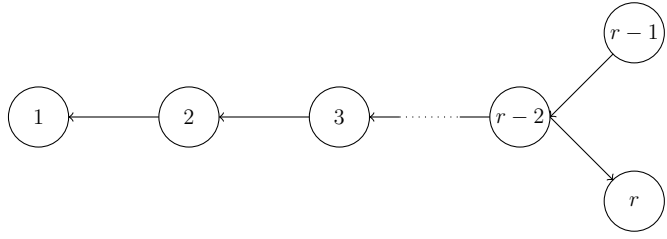

In this section, we consider 5d $\CN=1$ gauge theories based on the quiver $D_r$ Dynkin diagram (Figure \ref{fig_Dr}). The theory will be of pure gauge, i.e. we disregard the presence of matter fields in fundamental/anti-fundamental representation. Each node $s$ is associated to a gauge group U($m_s$) with exponentiated gauge coupling $\qf_s$, and a Chern-Simons term of coupling $\k_s$. The theory is considered on the Coulomb branch, and the exponentiated vacuum expectation values of the Higgs field form a vector of rank $m_s$ denoted $\vec v_s=(v_1^{(s)},\cdots,v_{m_s}^{(s)})$,  Each link $<ss'>$ represents a bifundamental matter field of mass $\mu_{s,s'}$ transforming in the fundamental representation of the two gauge groups lying at the extremities of the link. The corresponding instanton partition function reads
\begin{align}\label{def_Zinst_Dr}
\begin{split}
\Zinst[D_r]=\sum_{\{\vec\l_s\}}&\prod_{s=1}^r\qf_s^{|\vec\l_s|}\Zv(\vec v_{s},\vec\l_s)\ZCS(\k_s,\vec\l_s)\\
&\times\prod_{s=1}^{r-2}\Zbf(\vec v_{s+1},\vec\l_{s+1},\vec v_s,\vec\l_s|\mu_{s+1,s})\times\Zbf(\vec v_{r-2},\vec\l_{r-2},\vec v_{r},\vec\l_r|\mu_{r-2,r}).
\end{split}
\end{align}
The explicit expression of the contributions $\Zv$, $\ZCS$ and $\Zbf$ can be found in Appendix \ref{secB}.

The instanton partition function (\ref{def_Zinst_Dr}) is invariant under the following rescaling of the parameters:
\begin{equation}
\vec v_{s}\to\a_s\vec v_{s},\quad\qf_s\to\a_s^{-\k_s}\qf_s,\quad \mu_{s,s'}\to\dfrac{\a_s}{\a_{s'}}\mu_{s,s'}.
\end{equation} 
Choosing $\a_s/\a_{s+1}=\g\mu_{s+1,s}$ for $s=1,\cdots,r-2$ and $\a_r/\a_{r-1}=\g^2\mu_{r-1,r-2}\mu_{r-2,r}$, it is possible to set all the bifundamental masses $\mu_{s,s'}$ to the value $\g^{-1}$. This particular value of the bifundamental masses is known to simplify the algebraic presentation \cite{BFHMZ}, and we will make this choice here too.

There are two possible conventions for the bifundamental contributions in (\ref{def_Zinst_Dr}), depending on the orientation of the arrows on each link. Here we have chosen the arrows $j\to i$ for $i<j<r$, in contradistinction with the convention employed in our previous paper \cite{BFHMZ}.\footnote{In fact, the choice of edge orientation can be absorbed in the Chern-Simons term and the gauge coupling using the property
\begin{equation}
\Zbf(\vec v_1,\vec\l_1,\vec v_2,\vec\l_2|\g^{-1})=\dfrac{\prod_{l=1}^{m_1}(-\g v_l^{(1)})^{|\vec\l_2|}}{\prod_{l=1}^{m_2}(-\g v_l^{(2)})^{|\vec\l_1|}}\ZCS(m_2,\vec\l_1)\ZCS(-m_1,\vec\l_2)\Zbf(\vec v_2,\vec\l_2,\vec v_1,\vec\l_1|\g^{-1}).
\end{equation}} As a result, we will be able work with the boundary state $|\Omega,\a\rangle\rangle$ instead of the dual state $\langle\langle\Omega,\a|$. This orientation flip amounts to exchange the intertwiners $\Phi\leftrightarrow\Phi^\ast$.

\begin{figure}
\begin{center}
	\begin{tikzpicture}
	\draw (0,0)--(1,0);
	\draw (0,0)--(0,-1);
	\draw (0,-1)--(1,-1);
	\draw (0,0)--(-0.71,0.71);
	\draw (0,-1)--(-0.71,-1.71);
	\draw[dashed] (1,1)--(1,-2);
	\node at (1.7,-0.5) {${\bf ON}^0$};
	\draw [->] (2.5,-0.5)--(3,-0.5);
	\draw (3.5,0)--(2.79,0.71);
	\draw (3.5,0)--(3.5,-1);
	\draw (3.5,-1)--(2.79,-1.71);
	\draw (3.5,-1)--(4.7,-1);
	\draw (4.7,0)--(4.3,0);
	\draw (3.5,0)--(4.3,0);
	\draw (4.3,1)--(4.3,0);
	\draw (4.3,0)--(4.3,-0.95);
	\draw (4.3,-1.05)--(4.3,-1.8);
	\draw[dashed] (4.7,1)--(4.7,-1.8);
	\node at (5.5,-0.5) {${\bf ON}^-$};
	\draw[->] (6.5,-0.5)--(7,-0.5);
	\draw (7.5,0)--(6.79,0.71);
	\draw (7.5,0)--(7.5,-1);
	\draw (7.5,-1)--(6.79,-1.71);
	\draw (7.5,-1)--(9.7,-1);
	\draw (9.7,-0.8)--(9.3,-0.8);
	\draw (7.5,0)--(8.5,0);
	\draw (8.5,0)--(9.3,-0.8);
	\draw (9.3,-0.8)--(9.3,-0.95);
	\draw (9.3,-1.05)--(9.3,-1.8);
	\draw (8.5,0)--(8.5,1);
	\draw[dashed] (9.7,1)--(9.7,-1.8);
	\node at (10.5,-0.5) {${\bf ON}^-$};
	\end{tikzpicture}
\end{center}
\caption{Resolution of the orientifold, ${\bf ON}^0\rightarrow {\bf ON}^-+{\rm NS5}$, and the conifold singularity gives rise to the ``resolved" brane web proposed in \cite{HKLTY}. }
\label{fig_conifold}
\end{figure}
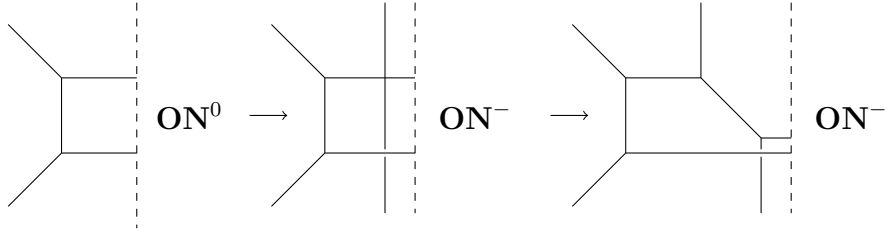

Our algebraic construction of the instanton partition function for D-type quiver gauge theory is inspired by the $(p,q)$-brane construction \cite{Kapustin-D,Hanany-Zaffaroni} employing an orientifold plane ${\bf ON}^0$ in order to reflect the D5 branes. However, instead of this ``macroscopic" picture, our method reveals itself closer from the ``resolved" picture of this orbifold construction which has been proposed in \cite{HKLTY}. This resolution is represented in the $(56)$-plane in Figure \ref{fig_conifold}.

\subsection{$D_1$ quiver}

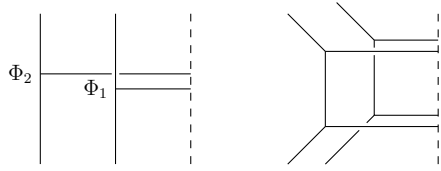
\begin{figure}
\begin{center}
\begin{tikzpicture}[scale=0.5]
\draw (0,2)--(0,6);
\draw (2,2)--(2,6);
\draw[dashed] (4,2)--(4,6);
\draw (2,4)--(4,4);
\draw (4,4.4)--(2.1,4.4);
\draw (1.9,4.4)--(0,4.4);
\node[scale=0.7,left] at (2,4) {$\Phi_1$};
\node[scale=0.7,left] at (0,4.4) {$\Phi_2$};
\end{tikzpicture}
\hspace{1cm}
\begin{tikzpicture}[scale=0.5]
\draw (0,0)--(1,1)--(1,3)--(0,4);
\draw (1,1)--(4,1);
\draw (1,3)--(4,3);
\draw (1,0)--(1.9,0.9);
\draw (2.1,1.1)--(2.3,1.3);
\draw (4,1.3)--(2.3,1.3)--(2.3,2.9);
\draw (4,3.3)--(2.3,3.3);
\draw (2.3,3.1)--(2.3,3.3)--(1.3,4.3);
\draw[dashed] (4,0)--(4,4);
\end{tikzpicture}
\caption{DIM representation web for a $D_1$ quiver (left) and the corresponding $(p,q)$-brane web diagram for a $U(2)$ gauge group (right).}
\label{fig_brane_D1}
\end{center}
\end{figure}

The Dynkin graphs $D_r$ of rank $r\leq3$ are isomorphic to (products) of linear Dynkin graphs:
\begin{equation}
D_1\simeq A_1,\quad D_2\simeq A_1\times A_1,\quad D_3\simeq A_3.
\end{equation} 
This fact implies certain relations between the instanton partition functions, namely
\begin{equation}
\Zinst[D_1]=\Zinst[A_1],\quad\Zinst[D_2]=\left(\Zinst[A_1])\right)^2,\quad \Zinst[D_3]=\Zinst[A_3].
\end{equation} 
Our brane construction has to reproduce these relations at small rank. We first focus on the case $r=1$ and examine the quantity
\begin{align}
\begin{split}
\CT[D_1]&=\Phi^{(n_1,m_1)}[u_1,\vec v_1]\otimes\Phi^{(n_2,m_2)}[u_2,\vec v_2]|\Omega,\a\rangle\rangle\\
&=\sum_{\vec\l}a_{\vec\l}\ \Phi^{(n_1,m_1)}_{\vec\l}[u_1,\vec v_1]\otimes\Phi^{(n_2,m_2)}_{\vec\l}[u_2,\vec v_2].
\end{split}
\end{align}
The projection on the boundary state imposes the constraints $m_1=m_2=m$ and $\vec v_1=\a^{-1}\vec v_2=\vec v$ on the vertical leg of the intertwiners. The corresponding brane structure is represented in Figure \ref{fig_brane_D1}.

Taking the vacuum expectation value in the horizontal spaces of the operator $\CT[D_1]$, we find
\begin{align}
\begin{split}
\la\CT[D_1]\ra&=\left(\langle\vac|\otimes\langle\vac|\right)\CT[D_1]\left(|\vac\rangle\otimes|\vac\rangle\right)\\
&=\sum_{\vec\l}a_{\vec\l}\ \langle\vac|\Phi^{(n_1,m_1)}_{\vec\l}[u_1,\vec v_1]|\vac\rangle\langle\vac|\Phi^{(n_2,m_2)}_{\vec\l}[u_2,\vec v_2]|\vac\rangle\\
&=\sum_{\vec\l}a_{\vec\l}\ t_{n_1,m_1}(\vec\l,u_1,\vec v_1)t_{n_2,m_2}(\vec\l,u_2,\vec v_2).
\end{split}
\end{align}
Using the expressions (\ref{def_a}) and (\ref{def_t}) of the coefficients $a_{\vec\l}$ and $t_{n,m}$, the vev $\la\CT[D_1]\ra$ reproduces the instanton partition function $\Zinst[D_1]=\Zinst[A_1]$ upon identification of gauge and Chern-Simons couplings,
\begin{equation}
\qf=u_1u_2\a^{-n_2}\g^{-\k}\prod_l(-v_l),\quad \k=-m-n_1-n_2.
\end{equation} 
The role of the boundary parameter $\a$ is marginal here. Comparing with the previous realization of the instanton partition function $\Zinst[A_1]$ recalled in section \ref{sec_web}, we observe that the reflection of the intertwiner $\Phi_2$ by the boundary states plays the same role as a dual intertwiner $\Phi_2^\ast$ with parameters
\begin{equation}
n_2^\ast=-m-n_2,\quad u_2^\ast=\dfrac{\a^{n_2}\g^{-m}}{u_2\prod_l(-v_l)}.
\end{equation} 

\subsection{$D_2$ quiver}
\begin{figure}
\begin{center}
\begin{tikzpicture}[scale=0.5]
\draw (0,0)--(0,6);
\draw (2,0)--(2,6);
\draw[dashed] (4,0)--(4,6);
\draw (2,4)--(4,4);
\draw (4,4.4)--(2.1,4.4);
\draw (1.9,4.4)--(0,4.4);
\draw (0,2)--(2,2);
\node[scale=0.7,left] at (2,4) {$\Phi_2$};
\node[scale=0.7,left] at (0,4.4) {$\Phi_4$};
\node[scale=0.7,left] at (0,2) {$\Phi_3$};
\node[scale=0.7,below left] at (2,2) {$\Phi_1^\ast$};
\end{tikzpicture}
\hspace{1cm}
\begin{tikzpicture}[scale=0.7,yscale=-1]
\draw (0,0)--(0,1);
\draw (0,0)--(-0.71,-0.71);
\draw (-0.71,-0.71)--(-2.71,-1);
\draw(0,0)--(2.8,0);
\draw(2.8,0.1)--(1,1);
\draw(-0.71,-0.71)--(3.45,-0.71);
\draw(2.8,0.1)--(3.51,-0.61);
\draw(3.51,-0.61)--(3.51,-1.5);
\draw(3.51,-0.61)--(4,-0.61);
\draw(3.55,-0.71)--(4,-0.71);
\draw(2.95,0)--(4,0);
\draw(2.8,0.1)--(4,0.1);
\draw[dashed] (4,3)--(4,-1.5);
\draw(0,1)--(1,1);
\draw(-0.71,1.71)--(0,1);
\draw(1,1)--(0.29,1.71);
\draw(-0.71,1.71)--(0.29,1.71);
\draw(0.29,1.71)--(0.29,2.71);
\draw(-0.71,1.71)--(-2.71,2.71);
\end{tikzpicture}
\caption{DIM representation web for a $D_2$ quiver (left) and the corresponding $(p,q)$-brane web diagram for gauge groups of rank $m_s=2$ (left).}
\label{fig_brane_D2}
\end{center}
\end{figure}
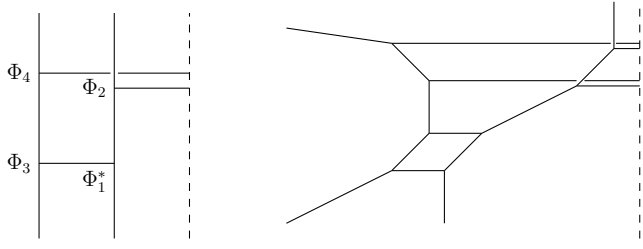

The operator associated to the $D_2$ quiver follows from the gluing rules given by the brane web diagram of Figure \ref{fig_brane_D2},
\begin{equation}
\bcontraction{\CT[D_2]=}{\Phi^{(n_1^\ast,m_2)\ast}[u_1,\vec v_1]}{\Phi^{(n_2,m_2)}[u_2,\vec v_2]\cdot}{\Phi^{(n_3,m_3)}[u_3,\vec v_3]}
\CT[D_2]=\Phi^{(n_1^\ast,m_2)\ast}[u_1,\vec v_1]\Phi^{(n_2,m_2)}[u_2,\vec v_2]\cdot\Phi^{(n_3,m_3)}[u_3,\vec v_3]\Phi^{(n_4,m_4)}[u_4,\vec v_4]\ |\Omega,\a\rangle\rangle.
\end{equation} 
Here the contraction symbol denotes a scalar product in the vertical space. The other two operators $\Phi_2$ and $\Phi_4$ are projected on the boundary state. The products $\Phi_1^\ast\Phi_2$ and $\Phi_3\Phi_4$ are products of operators in the Fock spaces. Expanded on the vertical components, the operator $\CT[D_2]$ reads
\begin{equation}
\CT[D_2]=\sum_{\vec\l_1,\vec \l_2}a_{\vec\l_1}a_{\vec\l_2}\ \Phi_{\vec\l_1}^{(n_1^\ast,m_1)\ast}[u_1^\ast,\vec v_1]\Phi_{\vec\l_2}^{(n_2,m_2)}[u_2,\vec v_2]\otimes\Phi_{\vec\l_1}^{(n_3,m_3)}[u_3,\vec v_3]\Phi_{\vec\l_2}^{(n_4,m_4)}[u_4,\vec v_4].
\end{equation} 
Gluing in vertical and horizontal channels leads to a number of constraints on the parameters:
\begin{align}\label{D2_constr}
\begin{split}
&m_1=m_3,\quad m_2=m_4,\quad n_3=n_4+m_4,\quad n_1^\ast+m_1=n_2+m_2,\\
&\vec v_1=\vec v_3,\quad \vec v_2=\a^{-1}\vec v_4,\quad u_3=u_4',\quad u_1^{\ast\prime}=u_2'.
\end{split}
\end{align}

We would like to show that the vev of the operator $\CT[D_2]$,
\begin{equation}
\la\CT[D_2]\ra=\sum_{\vec\l_1,\vec \l_2}a_{\vec\l_1}a_{\vec\l_2}\ \langle\vac|\Phi_{\vec\l_1}^{(n_1^\ast,m_1)\ast}[u_1^\ast,\vec v_1]\Phi_{\vec\l_2}^{(n_2,m_2)}[u_2,\vec v_2]|\vac\rangle\langle\vac|\Phi_{\vec\l_1}^{(n_3,m_3)}[u_3,\vec v_3]\Phi_{\vec\l_2}^{(n_4,m_4)}[u_4,\vec v_4]|\vac\rangle,
\end{equation} 
reproduces the instanton partition function of the $D_2$ quiver. The vev in each Fock space is easy to compute using the results (\ref{contraction_rule}) on the normal ordering of intertwiners:
\begin{align}
\begin{split}
&\langle\vac|\Phi_{\vec\l_1}^{(n_1^\ast,m_1)\ast}[u_1^\ast,\vec v_1]\Phi_{\vec\l_2}^{(n_2,m_2)}[u_2,\vec v_2]|\vac\rangle=\bZbf(\vec v_2,\vec \l_2,\vec v_1,\vec \l_1|\g^{-1})t_{n_1^\ast,m_1}^\ast(\vec\l_1,u_1^\ast,\vec v_1)t_{n_2,m_2}(\vec\l_2,u_2,\vec v_2)\\
&\langle\vac|\Phi_{\vec\l_1}^{(n_3,m_3)}[u_3,\vec v_3]\Phi_{\vec\l_2}^{(n_4,m_4)}[u_4,\vec v_4]|\vac\rangle=\bZbf(\a\vec v_2,\vec\lambda_2,\vec v_1,\vec\lambda_1|1)^{-1}t_{n_3,m_1}(\vec\l_1,u_3,\vec v_1)t_{n_4,m_2}(\vec\l_2,u_4,\a\vec v_2)
\end{split}
\end{align}
Note that we have used the constraints (\ref{D2_constr}) to eliminate the parameters $m_3,m_4,\vec v_3,\vec v_4$. The (rescaled) bifundamental contribution obeys the scaling property
\begin{equation}\label{scaling_Zbf}
\bZbf(\vec v_1,\vec\l_1,\vec v_2,\vec\l_2|\mu)= \bZbf(\mu^{-1}\vec v_1,\vec\l_1,\vec v_2,\vec\l_2|1)=\bZbf(\vec v_1,\vec\l_1,\mu\vec v_2,\vec\l_2|1).
\end{equation} 
Thus, we observe a cancellation of the bifundamental contributions for the specific value $\a=\g$ of the boundary parameter. In this case, the vev of the operator $\CT[D_2]$ takes the factorized form
\begin{equation}
\la\CT[D_2]\ra=\left(\sum_{\vec\l_1}a_{\vec\l_1}t_{n_1^\ast,m_1}^\ast(\vec\l_1,u_1^\ast,\vec v_1)t_{n_3,m_1}(\vec\l_1,u_3,\vec v_1)\right)\left(\sum_{\vec\l_2}a_{\vec\l_2}t_{n_2,m_2}(\vec\l_2,u_2,\vec v_2)t_{n_4,m_2}(\vec\l_2,u_4,\a\vec v_2)\right).
\end{equation} 
We recover here the product of two $A_1$ instanton partition functions,
\begin{equation}
\la\CT[D_2]\ra=\left(\sum_{\vec\l_1}\qf_1^{|\vec\l_1|}\Zv(\vec v_1,\vec\l_1)\ZCS(\k_1,\vec\l_1)\right)\left(\sum_{\vec\l_2}\qf_2^{|\vec\l_2|}\Zv(\vec v_2,\vec\l_2)\ZCS(\k_2,\vec\l_2)\right),
\end{equation} 
upon the identifications
\begin{equation}
\k_1=n_1^\ast-n_3,\quad \k_2=-m_2-n_2-n_4,\quad\qf_1=\g^{-\k_1-m_1}u_3/u_1^\ast,\quad \qf_2=u_2u_4\g^{-n_4-\k_2}\prod_{l=1}^{m_2}(-v_l^{(2)}).
\end{equation} 
It is important to remark that the constraints (\ref{D2_constr}) on the algebraic quantities bring no further constraints on the gauge theory parameters. The rank of the two gauge groups $U(m_1)$ and $U(m_2)$ are clearly independent in (\ref{D2_constr}), and so are the two sets of Coulomb branch vevs. On the other hand, the identities (\ref{D2_constr}) imply that $\k_1=n_2-n_4-m_1$, which is still independent from $\k_2=-m_2-n_2-n_4$. A similar argument holds for the gauge couplings $\qf_1$ and $\qf_2$.

\begin{figure}
\begin{center}
\begin{tikzpicture}
\draw (0,0)--(-0.71,0.71);
\node at (-0.5,1) {$t$};
\draw (0,0)--(0,-1);
\node at (-0.3,-0.3) {$q$};
\draw (0,-1)--(-0.71,-1.71);
\node at (-0.5,-1.2) {$q$};
\node at (0.3,-0.7) {$t$};
\draw (0,-1)--(2.2,-1);
\draw (0.95,-0.9)--(0.95,-1.1);
\draw (1.05,-0.9)--(1.05,-1.1);
\draw (2.2,-0.8)--(1.8,-0.8);
\draw (0,0)--(1,0);
\draw (0.45,0.1)--(0.45,-0.1);
\draw (0.55,0.1)--(0.55,-0.1);
\draw (1,0)--(1.8,-0.8);
\draw (1.8,-0.8)--(1.8,-0.95);
\draw (1.8,-1.05)--(1.8,-1.8);
\draw (1,0)--(1,1);
\draw[dashed] (2.2,1)--(2.2,-1.8);
\node at (1.2,0.5) {$t$};
\node at (1.5,-0.1) {$q$};
\node at (1.4,-0.7) {$t$};
\draw (1.95,-0.7)--(1.95,-0.9);
\draw (2.05,-0.7)--(2.05,-0.9);
\node at (3.8,-1) {$\frac{P'(\lambda^t;t,q)}{P'(\lambda;q,t)}f_\lambda(t,q)$};
\node at (1.6,-1.5) {$q$};
\end{tikzpicture}
\end{center}
\caption{Realization of the $D_2$ quiver with $U(1)\times U(1)$ gauge group by topological vertices and orientifold plane.}
\label{fig_D2_pq}
\end{figure}
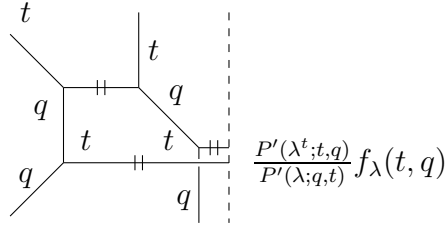

\paragraph{Remark} In the case of ranks $m_1=m_2=1$, our construction can be easily translated to the IKV topological vertex formalism (see Appendix \ref{secC} for a brief review). First we see that Figure \ref{fig_brane_D2} immediately reduces to the resolved brane web in Figure \ref{fig_conifold} in this case. The reflection state simply transposes $\Phi$ in the preferred direction (vertical representation). It thus becomes a ket, thereby allowing us to replace $\Phi^\ast$ by this object in the definition of the operator $\CT$. It is possible to reproduce the decoupling of the $U(1)$ components of the gauge group of $U(1)\times U(1)$ by assigning the factor $f_\lambda(t,q)P'(\lambda^t;t,q)/P'(\lambda;q,t)$ to the reflection state which is motivated by the replacement of $\Phi^\ast$ by $\Phi$ (see Figure \ref{fig_D2_pq}). 

Let us see how the cancellation occurs in the IKV formalism. According to the prescription given above, the instanton partition function of the $U(1)\times U(1)$ gauge theory reads 
\ba
\Zinst[D_2](Q_1,Q_2)= \sum_{\mu,\nu,\lambda,\sigma}Q_1^{|\lambda|}Q_2^{|\sigma|}Q^{|\mu|+|\nu|}\lt(\frac{q}{t}\rt)^{\frac{1}{2}|\nu|}f_\mu(t,q)f_\sigma(t,q)P'(\sigma^t;t,q)/P'(\sigma;q,t)\nn\\
\times C_{\emptyset\mu\lambda}(t,q)C_{\nu\emptyset\lambda^t}(q,t)C_{\mu^t\emptyset\sigma}(t,q)C_{\nu^t\emptyset\sigma}(t,q),
\ea
where the sum if taken over the realizations of the Young diagrams $\mu$, $\nu$, $\lambda$ and $\sigma$, and $Q_1$, $Q_2$, $Q$ are related to K\"ahler parameters of the local geometry\footnote{In terms of the AFS approach, $Q=v_1/v_2$, and $Q_{1,2}=\gamma^{-1}\mathfrak{q}_{1,2}$. }. We note that the factor $\lt(\frac{q}{t}\rt)^{\frac{1}{2}|\nu|}$ exactly corresponds to the $\gamma$-shift of the Coulomb branch parameter $v_2$ in the AFS approach. We can extract from this partition function the factors corresponding to the contribution of the horizontal representations in the AFS approach. There are two such factors\footnote{The factor $(-\sqrt{q/t})^{|\mu|}$ comes from a combination of several additional factors, including the framing factor, $f_\mu(t,q)\lt(\frac{q}{t}\rt)^{\frac{||\mu||^2}{2}}t^{-\frac{\kappa(\mu)}{2}}=(-\sqrt{q/t})^{|\mu|}$. The calculation is done using the following identities:
\begin{equation}
n(\nu)=\frac{1}{2}(||\nu^t||^2-|\nu^t|)=\sum_{x\in \nu}a(x),\quad n(\nu^t)=\frac{1}{2}(||\nu||^2-|\nu|)=\sum_{x\in \nu}\ell(x),\quad 
n(\nu)=n(\nu^t)-\frac{1}{2}\kappa(\nu^t)=n(\nu^t)+\frac{1}{2}\kappa(\nu).
\end{equation}
The functions $n(\l)$, $|\l|$, $||\l||$ and $\k(\l)$ of a Young diagram $\l$ are defined in the Appendix \ref{secC}.}, 
\ba
&&\sum_\mu (-Q\sqrt{q/t})^{|\mu|}s_\mu(t^{-\lambda^t}q^{-\rho})s_\mu(q^{-\sigma}t^{-\rho}),\nn\\
&&\sum_\nu (Q\sqrt{q/t})^{|\nu|}s_{\nu^t}(q^{-\lambda^t}t^{-\rho})s_\nu(q^{-\sigma}t^{-\rho}),\nn
\ea
which obviously cancel each other due to the following identities involving Schur functions, 
\ba
\sum_\mu s_\mu(x)s_\mu(y)=\prod_{i,j}(1-x_iy_j)^{-1},\quad \sum_\nu s_{\nu^t}(x)s_{\nu}(y)=\prod_{i,j}(1+x_iy_j).
\ea

It is also possible to generalize this prescription to $D_2$ quivers with higher-rank gauge groups using more general identities of similar type for Schur functions. These identities follow from the vertex-operator realization of the skew Schur function. 
\ba
s_{\mu/\eta}(\vec{x})=\bra{\eta}\tilde{V}_+(\vec{x})\ket{\mu}=\bra{\mu}\tilde{V}_-(\vec{x})\ket{\eta},\label{skew-schur}
\ea
where 
\ba
\tilde{V}_\pm (\vec{x})=\exp\lt(\sum_{n=1}^\infty \frac{1}{n}\sum_i x_i^n J_{\pm n}\rt),
\ea
and $J_n:=\sum_{j\in \mathbb{Z}+1/2}\psi_{-j}\psi^\ast_{j+n}$ denotes the modes of a free boson constructed from two Neveu-Schwarz free fermions, whose modes satisfy $\{\psi_n,\psi_m\}=\{\psi_n^\ast,\psi_m^\ast\}=0$ and $\{\psi_n,\psi^\ast_m\}=\delta_{n+m,0}$. The free fermion basis labeled by a Young diagram here takes the form $\ket{\mu}=\prod_{i=1}^d\psi_{-\alpha_i}\prod_{i=1}^d\psi^\ast_{-\beta_i}\ket{{\rm vac}}$, where $\mu=(\alpha_1,\alpha_2,\dots|\beta_1,\beta_2\dots)$ denotes the set of Frobenius coordinates of the Young diagram: $\a_i=\l_i-i$ and $\b_i=\l_i'-i$ with $i$ ranging from one to the number of squares on the diagonal. When all the external legs are trivial, factors corresponding to the horizontal representation can be written as the vev of vertex operators of a free boson, $\tilde{V}_\pm(\vec{x})$. The cancellation of bifundamental factors occurs essentially in the same way as in the AFS approach, and thus we omit the details here. 



\subsection{$D_r$ quiver}
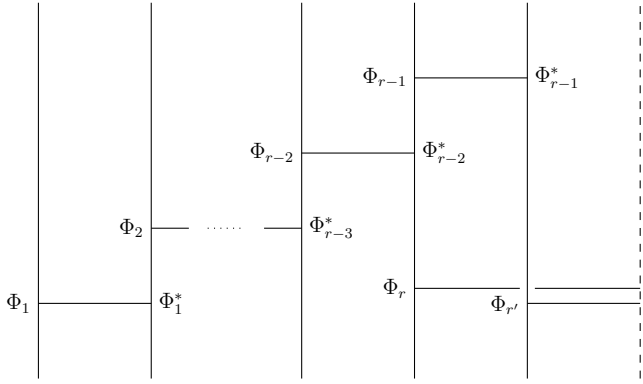
\begin{figure}
\begin{center}
\begin{tikzpicture}[scale=0.5]
\draw (0,0)--(0,10);
\draw (3,0)--(3,10);
\draw (7,0)--(7,10);
\draw (10,0)--(10,10);
\draw (13,0)--(13,10);
\draw[dashed] (16,0)--(16,10);
\draw (0,2)--(3,2);
\draw (3,4)--(4,4);
\draw[dotted] (4.5,4)--(5.5,4);
\draw (6,4)--(7,4);
\draw (7,6)--(10,6);
\draw (10,8)--(13,8);
\draw (10,2.4)--(12.8,2.4);
\draw (13.2,2.4)--(16,2.4);
\draw (13,2)--(16,2);
\node[scale=0.7,left] at (0,2) {$\Phi_1$};
\node[scale=0.7,right] at (3,2) {$\Phi_1^\ast$};
\node[scale=0.7,left] at (3,4) {$\Phi_2$};
\node[scale=0.7,right] at (7,4) {$\Phi_{r-3}^\ast$};
\node[scale=0.7,left] at (7,6) {$\Phi_{r-2}$};
\node[scale=0.7,left] at (10,8) {$\Phi_{r-1}$};
\node[scale=0.7,right] at (10,6) {$\Phi_{r-2}^\ast$};
\node[scale=0.7,left] at (10,2.4) {$\Phi_{r}$};
\node[scale=0.7,below left] at (13,2.4) {$\Phi_{r'}$};
\node[scale=0.7,right] at (13,8) {$\Phi_{r-1}^\ast$};
\end{tikzpicture}
\caption{DIM representation web for a general $D_r$ quiver}
\label{fig_brane_Dr}
\end{center}
\end{figure}

The construction of the general $D_r$ quiver gauge theory is obtained by combining the arguments of the previous section with the construction given in \cite{BFHMZ} for the $A_r$ quivers. For this purpose, we define the operator
\begin{equation}\label{def_CT_Dr}
\bcontraction{\CT[D_r]=}{\Phi^{(1)}}{\cdot}{\Phi^{(1)\ast}}
\bcontraction{\CT[D_r]=\Phi^{(1)}\cdot\Phi^{(1)\ast}\Phi^{(2)}\cdot\cdots\cdot\Phi^{(r-3)\ast}}{\Phi^{(r-2)}}{\cdot\Phi^{(r-1)}}{\Phi^{(r-2)\ast}}
\bcontraction{\CT[D_r]=\Phi^{(1)}\cdot\Phi^{(1)\ast}\Phi^{(2)}\cdot\cdots\cdot\Phi^{(r-3)\ast}\Phi^{(r-2)}\cdot\Phi^{(r)}\Phi^{(r-2)\ast}}{\Phi^{(r)}}{\cdot\Phi^{(r')}}{\Phi^{(r-1)\ast}}
\CT[D_r]=\Phi^{(1)}\cdot\Phi^{(1)\ast}\Phi^{(2)}\cdot\cdots\cdot\Phi^{(r-3)\ast}\Phi^{(r-2)}\cdot\Phi^{(r)}\Phi^{(r-2)\ast}\Phi^{(r-1)}\cdot\Phi^{(r')}\Phi^{(r-1)\ast}|\Omega,\a\rangle\rangle,
\end{equation} 
where we have employed the shortcut notations
\begin{equation}
\Phi^{(i)}=\Phi^{(n_i,m_i)}[u_i,\vec v_i],\quad\Phi^{(i)\ast}=\Phi^{(n_i^\ast,m_i)\ast}[u_i^\ast,\vec v_i].
\end{equation}
In (\ref{def_CT_Dr}), the contraction symbol is a reminder of which operators are glued in the vertical space using the scalar product. The two remaining intertwiners $\Phi^{(r)}$ and $\Phi^{(r')}$ are projected on the boundary state. Expanded over its vertical components, the operator $\CT[D_r]$ reads
\begin{equation}
\CT[D_r]=\sum_{\{\vec\l_s\}}\prod_{s=1}^r a_{\vec\l_s}\ \Phi_{\vec\l_1}^{(1)}\otimes\Phi_{\vec\l_1}^{(1)\ast}\Phi_{\vec\l_2}^{(2)}\otimes\cdots\otimes\Phi_{\vec\l_{r-3}}^{(r-3)\ast}\Phi_{\vec\l_{r-2}}^{(r-2)}\otimes\Phi_{\vec\l_r}^{(r)}\Phi_{\vec\l_{r-2}}^{(r-2)\ast}\Phi_{\vec\l_{r-1}}^{(r-1)}\otimes\Phi_{\vec\l_r}^{(r')}\Phi_{\vec\l_{r-1}}^{(r-1)\ast}.
\end{equation} 
This operator is well-defined provided that the algebraic parameters obey the following constraints ($i=1\cdots r-2$):
\begin{align}
\begin{split}
&m_r=m_{r'},\quad n_i^\ast+m_i=n_{i+1}+m_{i+1},\quad n_{r-2}^\ast=n_r,\quad n_{r-1}^\ast=n_{r'},\\
&\vec v_r=\a^{-1}\vec v_{r'},\quad u_i^{\ast\prime}=u_{i+1}',\quad u_{r-2}^\ast=u_{r},\quad u_{r-1}^{\ast}=u_{r'}
\end{split}
\end{align}

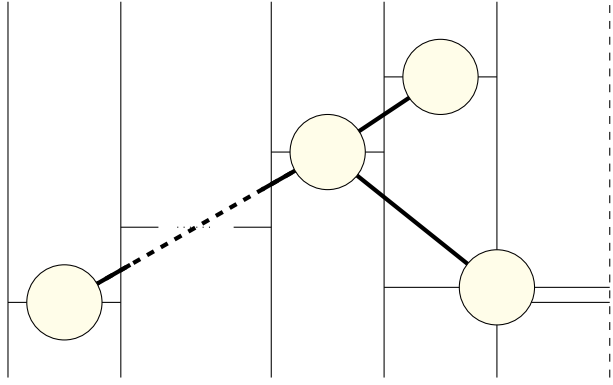
\begin{figure}
	\begin{center}
		\begin{tikzpicture}[scale=0.5]
		\draw (0,0)--(0,10);
		\draw (3,0)--(3,10);
		\draw (7,0)--(7,10);
		\draw (10,0)--(10,10);
		\draw (13,0)--(13,10);
		\draw[dashed] (16,0)--(16,10);
		\draw (0,2)--(3,2);
		\draw (3,4)--(4,4);
		\draw[dotted] (4.5,4)--(5.5,4);
		\draw (6,4)--(7,4);
		\draw (7,6)--(10,6);
		\draw (10,8)--(13,8);
		\draw (10,2.4)--(12.8,2.4);
		\draw (13.2,2.4)--(16,2.4);
		\draw (13,2)--(16,2);
		\draw[ultra thick, dashed] (1.5,2)--(8.5,6);
		\draw[ultra thick] (1.5,2)--(3.25,3);
		\draw[ultra thick] (6.75,5)--(8.5,6);
		\draw[ultra thick] (8.5,6)--(13,2.4);
		\draw[ultra thick] (11.5, 8)--(8.5,6);
		\filldraw [fill=yellow!10] (1.5,2) circle (1cm);
		\filldraw [fill=yellow!10] (8.5,6) circle (1cm);
		\filldraw [fill=yellow!10] (11.5,8) circle (1cm);
		\filldraw [fill=yellow!10] (13,2.4) circle (1cm);
		\end{tikzpicture}
		\caption{$D$-type quiver diagram in representation web }
		\label{fig_brane_Dr_dynkin}
	\end{center}
\end{figure}

The vev of the operator $\CT[D_r]$ is obtained along the same line as before. For the specific value $\a=\g$, we observe a similar cancellation of bifundamental contributions associated to the tail of the operator, namely
\begin{align}
\begin{split}
&\langle\vac|\Phi_{\vec\l_r}^{(r)}\Phi_{\vec\l_{r-2}}^{(r-2)\ast}\Phi_{\vec\l_{r-1}}^{(r-1)}|\vac\rangle=\bZbf(\vec v_{r-2},\vec\l_{r-2},\vec v_r,\vec\l_r|\g^{-1})\dfrac{\bZbf(\vec v_{r-1},\vec\l_{r-1},\vec v_{r-2},\vec\l_{r-2}|\g^{-1})}{\bZbf(\vec v_{r-1},\vec\l_{r-1},\vec v_r,\vec\l_r|1)}t_{r-1}t_{r-2}^\ast t_r\\
&\langle\vac|\Phi_{\vec\l_r}^{(r')}\Phi_{\vec\l_{r-1}}^{(r-1)\ast}|\vac\rangle=\bZbf(\vec v_{r-1},\vec\l_{r-1},\a\vec v_r,\vec\l_r|\g^{-1})t_{r-1}^\ast t_{r'}
\end{split}
\end{align}
where we have used abbreviated but straightforward notations. The cancellation is a consequence of the scaling property (\ref{scaling_Zbf}) of the bifundamental contribution. 
We illustrate the Dynkin diagram appearing in the representation web in Figure \ref{fig_brane_Dr_dynkin}. The disconnection between two right nodes is a consequence of such cancellation. 
Thus, at $\a=\g$, the vev of the operator $\CT[D_r]$ reproduces the $D_r$ quiver instanton partition function (\ref{def_Zinst_Dr}), up to a perturbative factor of $\CG$-functions,
\begin{equation}
\la\CT[D_r]\ra=\dfrac{\Zinst[D_r]}{\prod_{i\to j}\prod_{l_i=1}^{m_i}\prod_{l_j=1}^{m_j}\CG(v_{l_i}^{(i)}/\g v_{l_j}^{(j)})},
\end{equation} 
upon the following identification of the couplings ($i=1\cdots r-1$):
\begin{equation}
\k_i=n_i^\ast-n_i,\quad \k_r=-m_r-n_r-n_{r'},\quad\qf_i=\g^{-\k_i-m_i}u_i/u_i^\ast,\quad \qf_r=u_ru_{r'}\g^{-n_{r'}-\k_r}\prod_{l=1}^{m_r}(-v_l^{(r)}).
\end{equation} 

\section{Conclusion}
In this paper, we have introduced the notion of reflection states in the vertical and horizontal representations of Ding-Iohara-Miki algebra. These states are closely related to the involutive morphisms of algebras $\s_V$ and $\s_H$ that send DIM to $\overline{\mathrm{DIM}}$. In the $(p,q)$-brane construction, $\s_V$ and $\s_H$ are interpreted as reflection symmetries of the $(56)$-plane in the $5$ (D5) and $6$ (NS5) directions respectively.\footnote{For an interpretation of these symmetries from an  integrability perspective, see \cite{Bourgine2017,Bourgine2017a}.} The vertical reflection state has been employed to engineer the 5d $\CN=1$ $D_r$-quiver SYM, with unitary gauge groups, from the representation theory of DIM algebra. The instanton partition function of these gauge theories has been properly reproduced as the vacuum expectation value of an operator $\CT[D_r]$. This operator has been obtained as a product of intertwiners following the gluing rules prescribed by the $(p,q)$-web diagram. In this construction, the vertical reflection state is in correspondence with the $\textbf{ON}^-$ orientifold brane of the resolved web diagram \cite{HKLTY}.

Horizontal and vertical representations are related under the action of Miki's automorphism $\CS$. This automorphism is expected to be a manifestation of the $\CS$-duality in IIB string theory. It is then natural to expect the horizontal boundary state to play a role in the construction of $\CN=1$ $A_n$-quiver SYM theories with D-type (i.e. $SO(2m)$) gauge groups since such theories are known to be the $\CS$-dual of the D-type quiver gauge theories.

The construction of $E_r$-quiver gauge theory forms another interesting open problem. In this case, the brane construction is not known. However, the correspondence with DIM representations might provide a hint on the appropriate web diagram. We hope to come back to this problem in a near future.

In \cite{BFHMZ}, qq-characters \cite{NPS,Nekrasov2015,Nekrasov2016,Nekrasov2016b,Kim2016} of linear quivers have been re-derived from the DIM algebraic construction (see also \cite{BMZ,5dBMZ} for earlier work on qq-characters and quantum algebras). Specifically, qq-characters are obtained as the vev of the product $\CX\CT$ where $\CX$ is an operator commuting with $\CT$. In the present work, we have constructed the operators $\CT[D_r]$ relevant to $D_r$-quivers. By examination of their covariance properties under the action of DIM algebra, it should be possible to define the $\CX$-operators as well, hence reconstructing the $D_r$ qq-characters. Research along this direction is currently in progress.

\section*{Acknowledgement}
We would like to thank Antonio Sciarappa, Joonho Kim, Kimyeong Lee and Taro Kimura for discussions.
JEB, YM, RDZ thank the mathematical research institute MATRIX in Australia where part of this research was performed at the occasion of the workshop ``Integrability in Low Dimensional Quantum Systems''. YM is partially supported by Grants-in-Aid for Scientific Research (Kakenhi $\#$25400246) from MEXT, Japan. MF and RDZ are supported by JSPS fellowship for Young students. RZ would also like to thank the hospitality of KIAS and the workshop {\it Strings and Fields 2017} held by YITP, where he had the opportunity to meet one of the authors of \cite{Kim2017}.

\appendix

\section{Hopf algebraic structure}\label{AppA}
The DIM algebra has the structure of a Hopf algebra with the Drinfeld coproduct, 
\begin{align}
\begin{split}\label{AFS_coproduct}
&\Delta(x^+(z))=x^+(z)\otimes 1+\psi^-(\hg_{(1)}^{1/2}z)\otimes x^+(\hg_{(1)}z)\\
&\Delta(x^-(z))=x^-(\hg_{(2)} z)\otimes \psi^+(\hg_{(2)}^{1/2}z)+1\otimes x^-(z)\\
&\Delta(\psi^+(z))=\psi^+(\hg_{(2)}^{1/2}z)\otimes\psi^+(\hg_{(1)}^{-1/2}z)\\
&\Delta(\psi^-(z))=\psi^-(\hg_{(2)}^{-1/2}z)\otimes\psi^-(\hg_{(1)}^{1/2}z)\,.
\end{split}
\end{align}
The antipode can be easily deduced from this coproduct, but it will be of no use in our construction.

\section{Nekrasov partition function}\label{secB}

The bifundamental contribution with $U(m)\times U(m')$ gauge group can be decomposed as a product 
of \textit{Nekrasov factors},\footnote{The Nekrasov factors enjoy the property 
\begin{equation}
N(v_2,\l_2,v_1q_3^{-1},\l_1)=(-v_1)^{-|\l_2|}(-q_3v_2)^{|\l_1|}\prod_{x\in\l_1}\chi_x^{-1}\prod_{x\in\l_2}\chi_x\ N(v_1,\l_1,v_2,\l_2).
\end{equation}}
\begin{equation}\label{def_Zbf}
\Zbf(\vec v,\vec\l,\vec v',\vec\l'|\mu)=\prod_{l=1}^{m}\prod_{l'=1}^{m'}N(v_l,\l^{(l)},\mu v_{l'}',\l^{(l')\prime}).
\end{equation} 
Various expressions of the Nekrasov factors have been written, the one given here has been obtained by solving the discrete Ward identities derived in \cite{KMZ,BMZ},
\begin{equation}\label{def_nek_factor}
N(v_1,\l_1,v_2,\l_2)=\prod_{\superp{x\in\l_1}{y\in\l_2}}S\left(\dfrac{\chi_x}{\chi_y}\right)\times\prod_{x\in\l_1}\left(1-\dfrac{\chi_x}{q_3 v_2}\right)\times\prod_{x\in\l_2}\left(1-\dfrac{v_1}{\chi_x}\right),
\end{equation}
with
\begin{equation}
S(z)=\dfrac{(1-q_1z)(1-q_2z)}{(1-z)(1-q_3^{-1}z)},\quad g(z)=\dfrac{S(z)}{S(q_3z)}.
\end{equation} 

The vector multiplet contribution is expressed in terms of the Nekrasov factors as follows:
\begin{equation}
\label{Zbf}
\Zv(\vec v,\vec\l)=\prod_{l,l'=1}^m\dfrac1{N(v_l,\l^{(l)},v_{l'},\l^{(l')})}=\dfrac1{\Zbf(\vec v,\vec\l,\vec v,\vec\l|1)}.
\end{equation}
Finally, the Chern-Simons and fundamental/anti-fundamental contributions are expressed in terms of a simple product over all boxes in the Young diagrams,
\begin{equation}\label{ZCS}
\ZCS(\k,\vec\l)=\prod_{x\in\vec\l}\left(\chi_x\right)^\k,\quad \Zf(\vec \mu^{(\text{f})}_i,\vec \l_i)=\prod_{x\in\vec\l}\prod_{j=1}^{f_i}\left(1-\chi_xq_3^{-1}(\mu^{(\text{f})}_{i,j})^{-1}\right),\quad 
\Zaf(\vec \mu^{(\text{af})}_i,\vec \l_i)=\prod_{x\in\vec\l}\prod_{j=1}^{\tilde{f}_i}(1-\mu^{(\text{af})}_{i,j}\chi_x^{-1}).
\end{equation}

\section{Refined Topological Vertex}\label{secC}

In this Appendix, we give a brief review on the IKV refined topological vertex \cite{IKV}, which is an alternative refined approach to the partition function of the refined topological string, in contrast with the AFS vertex used in this paper. 

The refined topological vertex has three legs, and they are respectively referred to as the $t$-, $q$-directions and the preferred direction.  Each leg is labeled by a Young diagram (denoted with Greek letters) in the IKV formalism. 
\ba
C_{\mu\nu\lambda}(t,q)=\lt(\frac{q}{t}\rt)^{(||\nu||^2+||\lambda||^2)/2}t^{-\kappa(\nu)/2+||\lambda||^2/2}P'(\lambda;q,t)
\sum_{\eta}\lt(\frac{q}{t}\rt)^{\frac{|\eta|+|\mu|-|\nu|}{2}}
s_{\mu^t/\eta}(t^{-\rho}q^{-\lambda})s_{\nu/\eta}(t^{-\lambda^t}q^{-\rho}),\nn\\
\ea
where $s_{\mu/\nu}$ is the skew Schur function, 
\ba
P'(\lambda;q,t)=\prod_{(i,j)\in \lambda}\lt(1-t^{a(i,j)+1}q^{\ell(i,j)}\rt)^{-1},
\ea
and the notations involving Young diagrams are listed below:
\ba
|\lambda|=\sum_i\lambda_i,\quad a(i,j)=\lambda^t_j-i,\quad \ell(i,j)=\lambda_i-j,\\
||\lambda||^2=\sum_i\lambda_i^2\quad \kappa(\lambda)=2\sum_{(i,j)\in\lambda}(j-i),\\
t^{-\lambda}q^{-\rho}=\{t^{-\lambda_1}q^{1/2},t^{-\lambda_2}q^{3/2},t^{-\lambda_3}q^{5/2},\dots\},
\ea
where $\l_i$ denotes the $i$th column of the Young diagram $\l$, $\l^t$ is the transposed of $\l$, and $|\l|$ the number of boxes.

When gluing vertices, depending on the local geometry of the toric diagram, we sometimes also need to add an extra weight, which is usually called the framing factor \cite{Taki07}. The framing factor is given by $f_\lambda(t,q)^m$ in the case of a local geometry equivalent to $\cO(m-1)\oplus \cO(-m-1)\rightarrow \mathbb{CP}^1$, with
\ba
f_\lambda(t,q)=(-1)^{|\lambda|}t^{n(\lambda)}q^{-n(\lambda^t)},\quad n(\lambda):=\sum_i(i-1)\lambda_i.
\ea

\bibliography{Ddraft}

\end{document}